\documentclass[twocolumn,showpacs,preprintnumbers,amsmath,amssymb]{revtex4}
\usepackage{epsfig}
\usepackage{dcolumn}% Align table columns on decimal point
\usepackage{bm}% bold math
\DeclareGraphicsExtensions{.eps,.ps,.eps.gz,.ps.gz,.eps.Z,{}}

\newcommand{\vx}{{\bf x}}

\newcommand{\ve}{{\bf e}}

\newcommand{\vq}{{\bf q}}

\newcommand{\vu}{{\bf u}}

\newcommand{\vv}{{\bf v}}

\newcommand{\vw}{{\bf w}}
\newcommand{\vk}{{\bf k}}
\newcommand{\ii}{{\rm i}}
\newcommand{\dd}{{\rm d}}
\newcommand{\dr}{\partial}

\newcommand{\mF}{{\cal F}}
\newcommand{\mH}{{\cal H}}

\newcommand{\mP}{{\cal P}}

\newcommand{\mG}{{\cal G}}

\newcommand{\rhob}{\overline{\rho}}
\newcommand{\Dirac}{\delta_{\rm D}}

\newcommand{\beq}{\begin{equation}} 
\newcommand{\eeq}{\end{equation}} 
\newcommand{\beqa}{\begin{eqnarray}} 
\newcommand{\eeqa}{\end{eqnarray}} 
\newcommand{\bea}{\begin{array}} 
\newcommand{\ea}{\end{array}} 
\newcommand{\lag}{\langle} 
\newcommand{\rag}{\rangle}
\newcommand{\Om}{\Omega_{\rm m}}
\newcommand{\Tr}{{\rm Tr}}
\newcommand{\ha}{\hat{\alpha}}
\newcommand{\hb}{\hat{\beta}}
\newcommand{\hk}{\hat{\kappa}}
\newcommand{\ho}{\hat{\omega}}
\newcommand{\hG}{\hat{G}}
\newcommand{\hki}{\hat{\kappa}_{\infty}}
\newcommand{\hoi}{\hat{\omega}_{\infty}}
\newcommand{\vgamma}{\vec{\gamma}}
\newcommand{\vlambda}{\vec{\lambda}}
\newcommand{\vomega}{\vec{\omega}}

\begin{document}

\title{Propagators in Lagrangian space}

\author{Francis Bernardeau}
\affiliation{Service de Physique Th{\'e}orique,
         CEA/DSM/SPhT, Unit{\'e} de recherche associ{\'e}e au CNRS, CEA/Saclay,
         91191 Gif-sur-Yvette c{\'e}dex}
\affiliation{Canadian Institute for Theoretical Astrophysics, University of Toronto,
60 St. George street,
Toronto, Ontario M5S 3H8, Canada}         
\author{Patrick Valageas}
\affiliation{Service de Physique Th{\'e}orique,
         CEA/DSM/SPhT, Unit{\'e} de recherche associ{\'e}e au CNRS, CEA/Saclay,
         91191 Gif-sur-Yvette c{\'e}dex}
\vspace{.2 cm}
\date{\today}
\vspace{.2 cm}
\begin{abstract}

It has been found recently that propagators, e.g. the cross-correlation spectra of the
cosmic fields with the initial density field, decay exponentially at large-$k$ in an 
Eulerian description of the dynamics. We explore here similar quantities defined for 
a Lagrangian space description. We find that propagators in Lagrangian space do not 
exhibit the same properties: they are found not to be monotonic functions of
time, and to track back the linear growth rate at late time (but with a renormalized
amplitude). These results have been obtained 
with a novel method which we describe alongside. It allows the formal resummation
of the same set of diagrams as those that led to the known results in Eulerian space.
We provide a tentative explanation for the marked differences seen between the 
Eulerian and the Lagrangian cases, and we point out the role played by the vorticity
degrees of freedom that are specific to the Lagrangian formalism. This provides us 
with new insights into the late-time behavior of the propagators.

\end{abstract}

 \vskip2pc

\maketitle

\section{Introduction}

Although the details of the evolution of the large-scale structure of the 
universe are probably affected by the presence of baryonic matter, in the context 
of a dark matter dominated cosmic fluid, it is thought that the global properties of
the matter distribution at cosmological scales are essentially determined 
by those of a self-gravitating dust fluid.

A complete understanding of the development of gravitational instabilities in 
such a  fluid is still however an open problem. It is one of the central issues 
for the study of structure formation in observational cosmology and
this is for instance what pure $N$-body cosmological simulations attempt to solve. 
The Vlasov equation, that is the fluid limit of the Boltzmann
equation, entirely describes this system (see
\cite{1980lssu.book.....P} or \cite{2002PhR...367....1B} for details). 
This equation of motion applies to the so-called Eulerian description of the dynamics,
where the fluid properties are described through functions of fixed space-time 
coordinates (such as density and velocity fields). However, there exists an alternative 
description where the system is defined by the trajectories of particles, labelled
as a function of their initial positions. This is the Lagrangian formalism, which
takes advantage of the particle description of the fluid.
While this may not be a convenient description of the dynamics in
the fully developed nonlinear regime, it gives good insights of the dynamics in 
the early stages of the development of the gravitational clustering. The widely 
used Zel'dovich approximation \cite{1970A&A.....5...84Z} corresponds for instance 
to a description of the displacement field based on its linear approximation.

There exists a standard perturbative approach to study the development of 
gravitational instabilities
beyond the linear approximation. This approach, and the main results it led to, is 
described to a large extent in \cite{2002PhR...367....1B}. While it can can be useful 
for some specific observables, it fails to provide effective tools for describing 
the evolution of quantities such as the density power spectrum beyond the linear regime. 
Then, one still needs to use semi-analytic prescriptions. The ones that are mostly used
now, e.g. the so-called Peacock and Dodds formula \cite{1996MNRAS.280L..19P} or the 
Smith et al. formula \cite{2003MNRAS.341.1311S}, originate either from the near 
universal transform advocated in \cite{1991ApJ...374L...1H}, or are based on an even 
more empirical construction, the halo model (see \cite{2002PhR...372....1C}).
It is to be noted though that these prescriptions offer predictions for the power 
spectrum with relatively low accuracy, at the level of 10\%, and are insecure in cases 
of non standard cosmological models. Clearly there is a need to do better!

Recently there has been a revival of Perturbation Theory techniques (see
\cite{2006PhRvD..73f3519C, 2007A&A...465..725V, 2007astro.ph..3563M, 2007PhRvD..75d3514M} and also
\cite{2007arXiv0711.3407V} for an overview of these 
ideas). In particular the Renormalized Perturbation Theory (hereafter RPT) formalism 
introduced in \cite{2006PhRvD..73f3519C} suggests a new scheme
for the construction of perturbation theory expansions. It has
been successfully applied to the shape of the two-point propagator, 
\cite{2006PhRvD..73f3520C}, and consequently to the two-point density power spectrum 
\cite{2007arXiv0704.2783C}. One of the core objects of this approach are the so-called 
propagators. They can be viewed as the cross-correlation between the cosmic fluids 
(that can either be the local density contrast or the peculiar velocity divergence) 
and the initial density field. In particular, it has been found that these correlators 
decay exponentially in the large-$k$ limit (where $k$ is the
Fourier mode of interest). This result has been obtained analytically from a partial 
resummation of diagrams - in a perturbation theory point of view - that are thought 
to be the leading contributors of the high-$k$ behavior of this quantity. 
It has been furthermore confirmed in numerical simulations.

The aim of this paper is to consider similar quantities in the Lagrangian 
description of the dynamics. Thus, our goal is not to reconstruct the real 
space power spectrum from Lagrangian variables (as done in 
\cite{2007arXiv0711.2521M}) but to extend, to other
objects of interest that arise in the Lagrangian framework, exact PT results.
We first recall in section \ref{Lagrangian-approach} the basic
 ingredients of this description. To compute the high-$k$ limit of the propagators 
we then assume Gaussian initial conditions and that the same set of diagrams will provide
us with the leading contributions. To be more specific, those diagrams are those in 
which all loops are connected to the principal line. As shown 
in \cite{2007A&A..476..31V}, and explained in details here, this approximation amounts 
to linearize the motion equation for a mode evolution while the low-$k$ modes act as 
a random stochastic background. As we show in section \ref{2D-dynamics} for 
the 2D dynamics and in section \ref{3D-dynamics} for the 3D, although the modes of 
this stochastic background - assumed to be of Gaussian statistics - are in infinite 
number, their effects can be recast as those of a finite number of Gaussian random 
variables. This method reveals extremely powerful. We explicitly show the results it 
leads to for the 2D and 3D-Lagrangian propagators. We summarize in the last section 
what we have learned from these calculations.

\section{Lagrangian approach}
\label{Lagrangian-approach}

\subsection{Equations of motion}
\label{Equations-of-motion}

In Lagrangian approaches the global properties of the fluid are reconstructed from 
the individual particle trajectories, $\vx(\vq,t)$, labeled by their initial 
Lagrangian coordinate $\vq$. Thus, the Eulerian comoving position $\vx$ at time $t$ 
reads as
\beq
\vx=\vq+\Psi(\vq,t) ,
\label{displacement}
\eeq
where $\Psi(\vq,t)$ is the displacement field. Note that in 
Eq.(\ref{displacement}) we use the property that in standard cosmological 
scenarios the cold dark matter has a negligible initial velocity dispersion
(as opposed to ``hot'' dark matter scenarios). This allows us to fully define
the particles by their initial Lagrangian coordinate $\vq$ with a unique initial
peculiar velocity $\vv(\vq)$. Then, the equation of motion for each particle 
reads as, once the homogeneous 
expansion of the Universe has been taken into account, 
\beq
\frac{\dr^2\vx(\vq)}{\dr\tau^2}+\mH \frac{\dr\vx(\vq)}{\dr\tau} = -\nabla_{\vx} \phi(\vq) ,
\label{EulerLag}
\eeq
where $\tau=\int\dd t/a$ is the conformal time (and $a$ the scale factor) and
$\mH=\dd\ln a/\dd\tau$ the conformal expansion rate. The gravitational potential
$\phi$ is given by Poisson's equation
\beq
\Delta_{\vx} \phi= \frac{3}{2} \Om \mH^2 \delta(\vq) ,
\label{Poisson}
\eeq
where $\Om$ is the matter density cosmological parameter and 
$\delta(\vq)=(\rho-\rhob)/\rhob$ the matter density contrast. It is to be noted that 
in this expression the Laplacian is taken with respect to the $\vx$ coordinates while 
the fields are naturally given as a function of $\vq$ through the expression of the 
displacement field. We assume here that the density contrasts vanish at initial time. 
The conservation of matter then implies that
\beq
1+\delta(\vq) = \frac{1}{J(\vq)} \;\;\; \mbox{with} \;\;\; 
J(\vq)=\left|\det\left(\frac{\dr\vx}{\dr\vq}\right)\right| .
\label{Jacobian}
\eeq
Then, by taking the divergence with respect to the Eulerian coordinate $\vx$
of the equation of motion (\ref{EulerLag}) we obtain
\beq
J(\vq)\, \nabla_{\vx} . \left[ \frac{\dr^2\Psi(\vq)}{\dr\tau^2}
+\mH \frac{\dr\Psi(\vq)}{\dr\tau} \right]
= \frac{3}{2} \Om \mH^2 (J(\vq)-1)
\label{divEulerLag}
\eeq
where we used Poisson's equation. As in the Eulerian case, it is convenient to 
introduce the time coordinate $\eta$ and the function $f(\tau)$ defined from the 
linear growth rate $D_+(\tau)$ as, 
\beq
\eta=\ln D_+(\tau), \;\; 
f=\frac{\dd\ln D_+}{\dd\ln a}=\frac{\dd\ln D_+}{\mH \dd\tau} .
\label{eta}
\eeq
The linear growth rate $D_+(\tau)$ is the growing solution of
\beq
\frac{\dd^2 D_+}{\dd\tau^2}+\mH \frac{\dd D_+}{\dd\tau} = \frac{3}{2} \Om \mH^2 D_+ ,
\label{D+}
\eeq
which we normalize as $D_{+0}=1$ today. Then, Eq.(\ref{divEulerLag}) reads as
\beq
J(\vq)\, \nabla_{\vx} . \left[ \Psi''(\vq)+\left(\frac{3\Om}{2f^2}-1\right) 
\Psi'(\vq)\right] = \frac{3\Om}{2f^2} (J(\vq)-1)
\label{diveta}
\eeq
where we note with a prime the partial derivative with respect to time $\eta$.

In the following we will restrict the calculations to the Einstein-de Sitter case for 
which $\Om/f^2= 1$. It is to be noted however that for all models of cosmological 
interest we have  $\Om/f^2 \simeq 1$ so that this assumption is very mildly 
restrictive \cite{2002PhR...367....1B}. Thus, up to a good approximation, our results 
can be extended to $\Lambda$CDM cosmologies by substituting for the appropriate
linear growth rate $D_+(\tau)$. Then, the dependence on the cosmological parameters 
is fully contained in the time-redshift relation $\eta(z)$. 

Equation (\ref{diveta}) can be written in matrix form as
\beq
\Tr\left[ \left({\rm com}(\frac{\dr\vx}{\dr\vq})\right)^T . 
\left(\frac{\dr\Psi''}{\dr\vq}+\frac{1}{2} \frac{\dr\Psi'}{\dr\vq}\right)
\right] = \frac{3}{2} (J(\vq)-1) ,
\label{divmatrix}
\eeq
where ${\rm com}(\frac{\dr\vx}{\dr\vq})$ is the comatrix of $(\dr\vx/\dr\vq)$.
It is also given by:
\beq
\left(\frac{\dr\vq}{\dr\vx}\right) =\left(\frac{\dr\vx}{\dr\vq}\right)^{-1} 
= 
\left({\rm com}(\frac{\dr\vx}{\dr\vq})\right)^T /{\det(\frac{\dr\vx}{\dr\vq})} .
\label{comatrix}
\eeq
Thus, Eq.(\ref{divmatrix}) is the form of the equation of motion (\ref{EulerLag})
written in terms of the Lagrangian displacement field $\Psi$ alone. However, it is
not sufficient to fully determine the dynamics as can be noticed from the fact that
we only used the potential part of Eq.(\ref{EulerLag}) when we took the divergence in
Eq.(\ref{divEulerLag}). Thus, we must supplement Eq.(\ref{divmatrix}) with the 
rotational part:
\beq
\nabla_{\vx} \times \left[ \frac{\dr^2\Psi(\vq)}{\dr\tau^2}
+\mH \frac{\dr\Psi(\vq)}{\dr\tau} \right] = 0 .
\label{rotEulerLag}
\eeq
As is well-known from the Eulerian perturbation theory, the rotational part of the
Eulerian peculiar velocity field $\vv$ decays in the linear regime and a curl-free
initial velocity field remains potential to any order in perturbation theory 
\cite{2002PhR...367....1B,1980lssu.book.....P}
(but vorticity will be generated by shell-crossings, see \cite{1999A&A...343..663P} 
for an estimation of this effect). Then, one usually restricts the dynamics to the 
case of irrotational initial velocity fields, $\nabla_{\vx} \times\vv=0$,
so that Eq.(\ref{rotEulerLag}) simplifies to:
\beq
\nabla_{\vx} \times \Psi'(\vq) = 0 , \;\;\; \mbox{hence} \;\;\;
\frac{\dr\Psi'_i(\vq)}{\dr x_j} = \frac{\dr\Psi'_j(\vq)}{\dr x_i} ,
\label{rotPsi}
\eeq
which is of first order over time. In matrix form this constraint implies 
that~\cite{1994ApJ...427...51B},
\beq
\left(\frac{\dr\Psi'(\vq)}{\dr\vq}\right) . 
\left({\rm com}(\frac{\dr\vx}{\dr\vq})\right)^T 
\;\;\; \mbox{is a symmetric matrix} .
\label{rotmatrix}
\eeq

In three-dimensional space Eqs.(\ref{rotmatrix}) are cubic in $\Psi$ and $\Psi'$ 
(in general they are of the order of the number of space dimensions).
However, it is possible to derive equivalent equations that are quadratic in
$\Psi$ whatever the dimensionality of space. They can be obtained through the 
introduction of the velocity potential, $\Upsilon$, which the velocity field is 
assumed to derive from, $\psi'_{i}(\vq)\equiv \dr \Upsilon/\dr x_i$, in $\vx$ 
coordinates. Expressing $\Upsilon$ in term of $\Psi$ and imposing that
$\dr^2 \Upsilon/\dr q_i\dr q_{j}$ is symmetric leads to an equivalent set of equations 
of lower order in $\psi$, \cite{BernardeauBook}. These equations can also be derived 
explicitly from Eq.(\ref{rotPsi}) by multiplying it by 
$(\dr x_i/\dr q_m) (\dr x_j/\dr q_\ell)$,
\beq
\frac{\dr x_i}{\dr q_m} \frac{\dr\Psi'_i(\vq)}{\dr q_\ell} = 
\frac{\dr x_j}{\dr q_\ell} \frac{\dr\Psi'_j(\vq)}{\dr q_m} ,
\label{rotPsi2}
\eeq
a constraint that in matrix form states that,
\beq
\left(\frac{\dr\vx}{\dr\vq}\right)^T . \left(\frac{\dr\Psi'(\vq)}{\dr\vq}\right) 
\;\;\; \mbox{is a symmetric matrix}.
\label{rotmatrix2}
\eeq
Equations (\ref{rotPsi2})-(\ref{rotmatrix2}) are quadratic over $\Psi$ hence
they are more convenient to use than Eqs.(\ref{rotPsi})-(\ref{rotmatrix}) in three
(or more) dimensions~\footnote{Although these equations appear to be of different order
they are equivalent at least as long as the transform from $\vx$ to $\vq$ is regular,
i.e. before shell crossings. In 2D, both equations lead to the same constraint equation.
In 3D they lead to different sets of three equations that have the form of systems of 
linear equations in $\dr\psi'_{i}/\dr q_{j}$ ($\dr\psi_{i}/\dr q_{j}$ being treated 
as parameters). As such, one system can be transformed into the other by linear 
transforms.}.

\subsection{Linear regime}
\label{Linear-regime}

The first stages of the dynamics take place at a time when the deviations from 
the Hubble flow are small. Then, the equations of motion can be linearized over 
the displacement field $\Psi$. From Eq.(\ref{Jacobian}) the Jacobian $J(\vq)$ then 
reads up to linear order,
\beq
J_L(\vq)= 1+\Tr\left(\frac{\dr\Psi_L(\vq)}{\dr\vq}\right) 
= 1+\sum_i \Psi_{Li,i} = 1-\kappa_L ,
\label{JL}
\eeq
where we note with a subscript $L$ all linear quantities. Note also that hereafter 
we define $\Psi_{i,j}$ as the partial derivative of the displacement field with 
respect to Lagrangian coordinates,
\beq
\Psi_{i,j}(\vq) = \frac{\dr\Psi_i}{\dr q_j} ,
\label{Psiij}
\eeq
and we introduced its divergence $-\kappa$,
\beq
\kappa(\vq) =  -\nabla_{\vq} . \Psi(\vq) =  -\sum_i \frac{\dr\Psi_i(\vq)}{\dr q_i} .
\label{kappa}
\eeq
It is to be noted that at linear order $\kappa$ is nothing but the density contrast. 
Its time derivative is proportional to the velocity divergence. The motion equation 
(\ref{divmatrix}) naturally reads at linear order,
\beq
\kappa_L''(\vq)  + \frac{1}{2} \kappa_L'(\vq)  = \frac{3}{2} \kappa_L(\vq) ,
\label{kappaL}
\eeq
where we recover the two well-known growing and decaying linear modes:
\beq
\kappa_+ = e^{\eta} \;\;\; \mbox{and} \;\;\; \kappa_- = e^{-3\eta/2} .
\label{kappa+-}
\eeq

In the following, we shall assume that the initial conditions are such that only the 
linear growing mode is present (but it would be possible to set different initial 
conditions):
\beq
\kappa_L(\vq,\eta)  = e^{\eta} \kappa_{0}(\vq)  \;\;\; \mbox{hence} \;\;\; 
\delta_L(\vq,\eta) = e^{\eta} \kappa_{0}(\vq)  .
\label{kappa0}
\eeq
Note then that at this order the constraint, Eq.(\ref{rotmatrix}), implies that  
$\Psi_L'(\vq)$ is curl-free in $\vq$ coordinates, 
$\nabla_{\vq} \times \Psi_L'(\vq) = 0$, and so is the linear
displacement field. It is then entirely determined by its divergence $\kappa$.

\subsection{Correlators and propagators}
\label{Correlators-and-propagators}

Because of the mathematical structure of the theory, it is obviously very convenient 
to rewrite the motion equations in Fourier space. The Fourier components of the field 
are defined as,
\beq
\kappa(\vk) = \int\frac{\dd^n\vq}{(2\pi)^n} e^{-\ii\vk.\vq} \kappa(\vq) ,
\label{kappak}
\eeq
where $\dd^n\vq$ is the $n-$dimensional volume element. The Fourier components of the 
linear displacement field can be easily written in terms of the Fourier modes of the 
divergence field,
\beq
\Psi_L(\vk) = \ii \frac{\vk}{k^2} \kappa_L(\vk) .
\label{PsiL}
\eeq

Note that because of the assumed statistical homogeneity and isotropy of space, 
ensemble average of products of two Fourier modes vanish for modes that do not sum 
to zero. This property holds for equal as well as unequal time correlators. 
In the following, we furthermore consider Gaussian initial conditions. As it will 
turn out, this is a crucial property. It indeed determines the diagrammatic structure 
and the contributions to the quantities of interest. Within this assumption the 
entire statistical properties of the initial density field are defined by its power 
spectrum, $P_{0}$(k), such that:
\beq
\lag \kappa_{0}(\vk_1) \kappa_{0}(\vk_2) \rag = \Dirac(\vk_1+\vk_2) P_{0}(k_1) ,
\label{P0}
\eeq
where $\lag\,.\,\rag$ represents ensemble averages over the statistical process at 
the origin of the large-scale structure.

If the notion of power spectrum has been widely used in theoretical and observational 
cosmology since the early eighties, the notion of propagator is relatively new. It 
has been introduced in \cite{2006PhRvD..73f3519C} (see also \cite{2004A&A...421...23V}
for the more general notion of response functions). By definition it represents the 
ensemble average of the functional derivative of a given cosmic field component with 
respect to the initial field value. What we will be interested in here is the propagator 
between an initial convergence mode $\kappa_{0}(\vk)$ and the final convergence mode 
$\kappa(\vk',\eta)$ (the one defined with respect to the rotational parts vanishes for 
parity reasons in case of rotational-free initial conditions).
As $\kappa(\vk',\eta)$ is the result of a complex nonlinear process, it is formally 
a functional of the whole set of the initial density modes (only in the linear regime 
does it only depend on the same $\vk$ mode). We can then introduce the functional 
derivative of $\kappa(\vk',\eta)$ with respect to $\kappa_{0}(\vk)$:
$\partial \kappa(\vk',\eta)/ \partial\kappa_{0}(\vk)$. This is a stochastic quantity 
whose ensemble average does not vanish for $\vk=\vk'$. It defines the 
{\sl propagator}~\footnote{The propagator can also be obtained as the ensemble average 
of the product of $\kappa(\vk,\eta)$ and $\kappa_{0}(\vk')$,
$\lag \kappa(\vk,\eta) \kappa_{0}(\vk')  \rag  = \Dirac(\vk-\vk') G(k,\eta) P_{0}(k)$. 
As shown in \cite{2006PhRvD..73f3519C} the two quantities are identical for Gaussian 
initial conditions.},
\begin{equation}
\lag\frac{\partial \kappa(\vk',\eta)}{\partial\kappa_{0}(\vk)}\rag =
\Dirac(\vk-\vk') G(k,\eta) .
\label{Gdef}
\end{equation}
The goal of this paper is precisely to  investigate the behavior of the propagator
$G(k,\eta)$. In the linear regime the functional $\kappa[\kappa_{0}]$ is trivial and
given by Eq.(\ref{kappa0}) which implies that,
\beq
G_L(k,\eta) = e^{\eta} .
\label{GL}
\eeq
From a perturbation theory point of view, the functional $\kappa[\kappa_{0}]$
can be expanded in terms of the initial convergence field,
\begin{eqnarray}
\kappa(\vk,\eta)=\sum_{p=1}^{\infty}
\int\dd^n\vw_{1}\dots\dd^{n}\vw_{p}\,\Dirac\left(\vk-\sum_{i=1}^p\vw_{i}\right)
\nonumber\\
\times\mF^{(p)}(\vw_{1},\dots,\vw_{n};\eta) \, 
\kappa_{0}(\vw_{1})\dots\kappa_{0}(\vw_{p})
\label{kappaExpansion}
\end{eqnarray}
where the kernels $\mF^{(p)}$ are {\sl symmetric} functions of wave modes. They
are determined by the motion equations for $\kappa$ and $\omega$. Then we have
\begin{eqnarray}
\lefteqn{ G(k,\eta) =\sum_{p} \int\dd^n\vw_{1}\dots\dd^{n}\vw_{p-1} } \nonumber\\
&& \hspace{-1cm} \times p \mF^{(p)}(\vw_{1},\dots,\vw_{p-1},\vk;\eta)
\,\lag \kappa_{0}(\vw_{1})\dots\kappa_{0}(\vw_{p-1})\rag
\label{GExpansion}
\end{eqnarray}
(the ensemble average of the r.h.s. of this equation ensures that 
$\sum_{i=1}^{p-1}\vw_{i}=0$ so that $\Dirac(\vk-\sum_{i=1}^p\vw_{i})$ is transformed 
into $\Dirac(\vk-\vk')$). Such an expansion can be represented in a diagrammatic 
way by taking advantage  of the Gaussian initial conditions. This can serve as a basis
for resummation schemes. We shall illustrate this construction for the 2D Lagrangian 
dynamics first.

\section{2D dynamics}
\label{2D-dynamics}

The aim of this section is to derive explicitly the motion equations for the 
Lagrangian 2D dynamics and to explore the resulting propagator properties. 
Since its mathematical structure is simpler than for the 3D case, it serves to 
illustrate the method we develop here to compute the propagators.

\subsection{Decomposition over curl-free and divergence-less parts}
\label{Decomposition}

We investigate in this section the simpler case of a two-dimensional dynamics.
This corresponds to perturbations with $\Psi_3=0$ that do not depend on the third 
coordinate, $q_3$ or $x_3$. Therefore, the nonlinear dynamics is restricted to the
plane $(\ve_1,\ve_2)$ and particles exactly follow the Hubble expansion along the third 
axis $\ve_3$.
Then, it is convenient to decompose the Lagrangian displacement field over a curl-free
part $\chi$ and a divergence-less part $\lambda$ as
\beq
\Psi = \left( \bea{c} \Psi_1 \\ \Psi_2 \\ 0 \ea \right) = 
\left( \bea{c} \frac{\dr\chi}{\dr q_1}+\frac{\dr\lambda}{\dr q_2} \\
\frac{\dr\chi}{\dr q_2}-\frac{\dr\lambda}{\dr q_1} \\ 0 \ea \right) 
= \nabla_{\vq} . \chi + \nabla_{\vq} \times (\lambda \, \ve_3)  .
\label{chilambda}
\eeq
Here and in the following we note $\times$ the 3-dimensional vector product.
Then, the divergence $-\kappa$ reads
\beq
\kappa= - \nabla_{\vq}^2 \chi , \;\;\; \kappa(\vk) = k^2 \chi(\vk) .
\label{kappachi}
\eeq
In a similar fashion, we define the vorticity as
\beq
\omega = - \nabla_{\vq}^2 \lambda , \;\;\; \omega(\vk) = k^2 \lambda(\vk) .
\label{omegalambda}
\eeq
Then, the equation of motion (\ref{divmatrix}) reads in Fourier space as
%\begin{widetext}
\beqa
\lefteqn{\kappa''+\frac{1}{2}\kappa'-\frac{3}{2}\kappa  = 
\int \dd\vk_1\dd\vk_2\, \Dirac(\vk_1\!+\!\vk_2\!-\!\vk) }\nonumber\\
&& \hspace{-.4cm} \times\left\{ \! \alpha(\vk_1,\vk_2)\! 
\left[ \kappa_1 (\kappa_2''\!+\!\frac{1}{2}\kappa_2'\!-\!\frac{3}{4}\kappa_2) 
\!+\! \omega_1 (\omega_2''\!+\!\frac{1}{2}\omega_2'\!-\!\frac{3}{4}\omega_2) \right] 
\right.\nonumber \\
&& \hspace{-.3cm}\left. + \beta(\vk_1,\vk_2)\!
\left[ \omega_1 (\kappa_2''\!+\!\frac{1}{2}\kappa_2') 
\!-\! \kappa_1 (\omega_2''\!+\!\frac{1}{2}\omega_2')\!
+\!\frac{3}{2} \kappa_1 \omega_2 \right] \right\}
\label{divkappaomega}
\eeqa
%\end{widetext}
where we noted $\kappa_i=\kappa(\vk_i), \omega_i=\omega(\vk_i)$, and
we introduced the symmetric kernels
\beqa
\alpha(\vk_1,\vk_2) &=& \frac{\det(\vk_1,\vk_2)^2}{k_1^2k_2^2} , 
\label{defalpha} \\ 
\beta(\vk_1,\vk_2) &=& \frac{(\vk_1.\vk_2)\det(\vk_1,\vk_2)}{k_1^2k_2^2} ,
\label{defbeta}
\eeqa
with
\beq
\det(\vk_1,\vk_2) = k_{1,1} k_{2,2}-k_{1,2} k_{2,1} = \ve_3 . (\vk_1\times\vk_2).
\label{detk1k2}
\eeq
It is to be noted that, unlike their Eulerian counterparts, these kernels only depend 
on the relative angle between the wave modes.

Equation (\ref{divkappaomega}) can be written in integral form by using the Green's 
function $\mG(\eta,\eta')$ that is solution of
\beq
\left(\frac{\dd^2}{\dd\eta^2}+\frac{1}{2}\frac{\dd}{\dd\eta}-\frac{3}{2}\right)
\mG(\eta,\eta') = \Dirac(\eta-\eta'). 
\label{Green}
\eeq
It reads as
\beq
\mG(\eta,\eta') = \theta(\eta-\eta') \frac{2}{5} 
\left[ e^{(\eta-\eta')} - e^{-3(\eta-\eta')/2}\right] ,
\label{Greenretarded}
\eeq
where $\theta(\eta-\eta')$ is the Heaviside factor which enforces causality. 
This constraint fully determines $\mG(\eta,\eta')$ (whereas Eq.(\ref{Green}) alone 
does not select between advanced and retarded propagators or combinations of both).
Of course, in Eq.(\ref{Greenretarded}) we recognize the two linear modes of
Eq.(\ref{kappa+-}). Thus, we can write the solution of Eq.(\ref{divkappaomega})
as:
%\begin{widetext}
\beqa
\lefteqn{ \kappa  =  \kappa_L + \int_{-\infty}^{\eta} \dd\eta' \mG(\eta,\eta') 
\int \dd\vk_1\dd\vk_2\ \Dirac(\vk_1\!+\!\vk_2\!-\!\vk) } \nonumber\\
&& \hspace{-0.4cm} \times \left\{ \! \alpha(\vk_1,\vk_2)\! 
\left[ \kappa_1 (\kappa_2''\!+\!\frac{1}{2}\kappa_2'\!-\!\frac{3}{4}\kappa_2) 
\!+\! \omega_1 (\omega_2''\!+\!\frac{1}{2}\omega_2'\!-\!\frac{3}{4}\omega_2) \right] 
\right. \nonumber \\
&& \hspace{-0.2cm} \left. + \beta(\vk_1,\vk_2)\!
\left[ \omega_1 (\kappa_2''\!+\!\frac{1}{2}\kappa_2') 
\!-\! \kappa_1 (\omega_2''\!+\!\frac{1}{2}\omega_2')\!
+\!\frac{3}{2} \kappa_1 \omega_2 \right] \!\right\}
\label{intdivkappaomega}
\eeqa
%\end{widetext}
where all terms in the brackets are taken at time $\eta'$ in the past.

On the other hand, the curl-free Eulerian velocity constraint (\ref{rotmatrix}) reads
as
\beqa
\omega' \!\! & \! = \! & \!\! \int \dd\vk_1\dd\vk_2 \Dirac(\vk_1+\vk_2-\vk)
\left\{ \alpha(\vk_1,\vk_2) [\kappa_1 \omega_2'-\omega_1 \kappa_2'] 
\right.\nonumber \\
&& \left. +  \beta(\vk_1,\vk_2)
\left[ \kappa_1 \kappa_2' + \omega_1 \omega_2' \right] \right\} .
\label{rotkappaomega} 
\eeqa
From Sect.~\ref{Linear-regime} we can see that the linear vorticity vanishes,
$\omega_L=0$, and Eq.(\ref{rotkappaomega}) can be integrated as
\beqa
\omega & = & \int_{-\infty}^{\eta} \dd\eta' \int \dd\vk_1\dd\vk_2 
\Dirac(\vk_1+\vk_2-\vk) \left\{ \alpha(\vk_1,\vk_2) \right. \nonumber \\
&& \hspace{-0.3cm} \left. \times [\kappa_1 \omega_2'-\omega_1 \kappa_2'] 
+ \beta(\vk_1,\vk_2) \left[ \kappa_1 \kappa_2' + \omega_1 \omega_2' \right] 
\right\} .
\label{introtkappaomega} 
\eeqa
In Eqs.(\ref{intdivkappaomega}) and (\ref{introtkappaomega}) we have set up the
initial conditions at time  $\eta_I\rightarrow-\infty$. It would be possible to keep 
$\eta_I$ finite, but this introduces extra terms in the perturbative series for 
$\kappa$ and $\omega$ that involve the decaying mode $\kappa_-$ of 
Eq.(\ref{kappa+-}). By contrast, from Eqs.(\ref{intdivkappaomega}), 
(\ref{introtkappaomega}), the nonlinear quantities
$\kappa$ and $\omega$ can be written as a perturbative series over powers of
the linear growing mode $e^{\eta}\,\kappa_{0}$, such that the term of order $p$
factorizes as $e^{p\eta}\,\kappa^{(p)}(\vk)$, as in the standard perturbation theory.

The kernels $\alpha(\vk_1,\vk_2)$ and $\beta(\vk_1,\vk_2)$ obey the symmetries
\beq
\alpha(\vk_1,\vk_2) = \alpha(\vk_2,\vk_1) , \;\; 
\beta(\vk_1,\vk_2)= - \beta(\vk_2,\vk_1) ,
\label{absym}
\eeq
as seen from Eqs.(\ref{defalpha})-(\ref{detk1k2}). This is consistent with the
fact that $\kappa$ and $\chi$ are scalars whereas $\omega$ and $\lambda$ are 
pseudoscalars, as seen from Eq.(\ref{chilambda}) 
(so that $\nabla_{\vq} \times (\lambda \, \ve_3)$
is a vector like $\Psi$). Then, under parity $\mP$ we have:
\beq
\mP : \kappa \rightarrow \kappa, \; \omega \rightarrow -\omega, 
\; \alpha \rightarrow \alpha , \; \beta \rightarrow -\beta .
\label{parity}
\eeq
Equation (\ref{parity}) actually implies that the kernels $\alpha,\beta$, satisfy
Eq.(\ref{absym}), since the exchange of basis vectors $\ve_1 \leftrightarrow \ve_2$
can be written as a rotation followed by a reflection. Then, the symmetry
(\ref{parity}) directly determines which kernel $\alpha$ or $\beta$ is associated
with a factor such as $\kappa\kappa$ or $\kappa\omega$ in Eqs.(\ref{divkappaomega})
and (\ref{rotkappaomega}).

\subsection{Diagrammatic representation}

\begin{figure*}
\centerline{\epsfig {figure=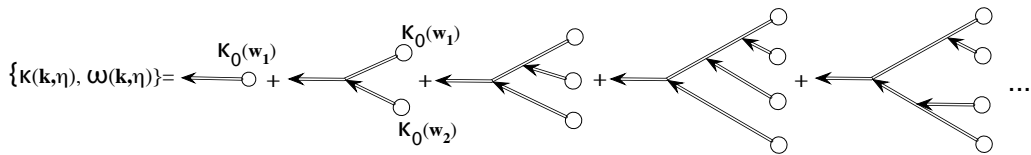,width=14cm}}
\caption{Diagrammatic expression of the expansion of the convergence-vorticity
doublet, $\{\kappa(\vk,\eta),\omega(\vk,\eta)\}$, in 2D dynamics. See text for details.} 
\label{KappaExpansion}
\end{figure*}

The equations (\ref{intdivkappaomega}, \ref{introtkappaomega}) have a simple 
diagrammatic representation which illustrates the fact that the functions $\mF^{(p)}$ 
are obtained from successive quadratic interactions. A diagrammatic expansion of 
equations (\ref{intdivkappaomega}, \ref{introtkappaomega}) is presented in 
Fig.~\ref{KappaExpansion}. Each open circle stands for a linear growing mode doublet
$\{\kappa_L,\omega_L\}=\{e^{\eta_j}\kappa_0(\vk_j),0\}$, whereas the vertex points 
represent the interaction operators that can be read out from 
Eqs.(\ref{intdivkappaomega}, \ref{introtkappaomega}).
For instance, its first component (the one that represents 
$\{\kappa(\vk_{1},\eta_{1}),\ \kappa(\vk_{2},\eta_{2})\} \to \kappa(\vk,\eta)$) is
\begin{eqnarray}
\lefteqn{ \hspace{-0.6cm} \gamma_{111}(\vk,\eta;\vk_1,\eta_1,\vk_2,\eta_2) 
= \int \dd\eta' \mG(\eta,\eta') \Dirac(\vk-\vk_1-\vk_2) }\nonumber\\
&& \times \Dirac(\eta_1-\eta')\Dirac(\eta_2-\eta')\,
\left(\frac{\partial^2}{\partial\eta_{2}^2}+\frac{1}{2}\frac{\partial}{\partial\eta_{2}}
-\frac{3}{4}\right) .
\label{gamma111}
\end{eqnarray} 
Then, one must integrate over the coordinates $(\vk_j,\eta_j)$ of the incoming modes
at each vertex.

\begin{figure*}
\centerline{\epsfig {figure=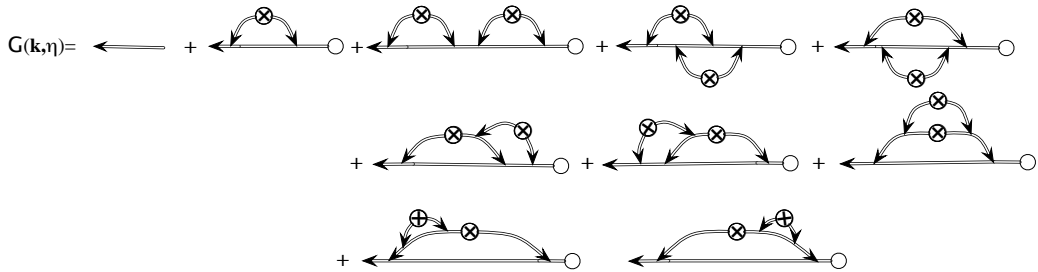,width=14cm}}
\caption{Diagrammatic expression of the expansion of the propagator $G(k,\eta)$ in 2D 
dynamics. All the contribution up to 2 loops are included. Note that the last two rows 
correspond to loop configurations that do not all connect to the "principal line" (shown 
here as a straight horizontal line). These are the contribution we assume to be 
subdominant. See text for details.}
\label{Gexpansion}
\end{figure*}

As noted in \cite{2006PhRvD..73f3520C}, each of these diagrams exhibits one and only one
``principal line'': a line that runs from the initial time to the final time without 
crossing a circle. It is then possible to sort the loop terms with respect to the number 
of vertices that are attached to this line. It is then expected that in the high-$k$ 
limit the dominant contribution comes from the diagrams whose number of such vertices 
are larger (see \cite{2006PhRvD..73f3520C} for details).
As seen in \cite{2007A&A..476..31V} and recalled below, this can be justified in a 
certain regime if it is possible to have a large separation of scales. 
In the following, we will restrict our calculations to this subset of contributions. 
For instance, for the  terms up to two-loop order, they correspond to the first row 
of Fig.~\ref{Gexpansion}.
It is important to note that these diagrams are such that the incoming waves are 
always in the linear regime. 
In \cite{2006PhRvD..73f3520C}, the authors were able 
to resum these loop contributions (basically by properly counting them). In a 
Lagrangian description things are made more difficult because of the complex nature 
of the vertices and it is not always possible to obtain an explicit analytical formula
for this resummed propagator.

However, as shown in \cite{2007A&A..476..31V}, the propagator defined by this partial 
series of diagrams can be seen as the exact propagator of a simpler dynamics (that only
gives rise to these diagrams). The latter can be derived by linearizing the equations
of motion in a certain fashion. Then, we can compute the propagator $G(k,\eta)$ by 
solving exactly this second dynamics and next performing the average over the initial 
conditions. This can be made numerically without performing diagrammatic resummations.
We first illustrate this alternative method for the 2D-Lagrangian dynamics.

\subsection{High-$k$ approximation}
\label{High-k}
%
%\begin{figure}
%\centerline{\epsfig {figure=HighkLinearProp.eps,width=8cm}}
%\caption{The double line represent the doublet $\{\kappa(\vk,\eta),\omega(\vk,\eta)\}$. Incoming
%bundle waves can be of type $\hat\alpha$ or $\hat\beta$.  The $\hat\alpha$ interactions preserve
%parity; they change the amplitude of the $\kappa$ and $\omega$ modes independently. The $\hat\beta$
%interactions change parity; they interchange $\kappa$ and $\omega$. For the ensemble average of
%this diagram not to vanish one needs to have an even number of both $\hat\alpha$ and $\hat\beta$
%incoming waves.}
%\label{HighkLinearProp}
%\end{figure}

\subsubsection{Resummation of dominant diagrams}
\label{Resummation}

As stated above, the dominant diagrams are expected to be those where all incoming 
lines to the principal path are in the linear regime. Following \cite{2007A&A..476..31V},
such a system is described by motion equations similar to (\ref{intdivkappaomega}) 
and (\ref{introtkappaomega}), where in the terms in the right hand sides 
we replace all terms except one by their linear values $\kappa_L$ and $\omega_L$ (the
latter vanishes here) and sum over all possible choices.
These equations correspond to a physical system where it is legitimate to separate
scales, for instance if there exists an upper wavenumber $\Lambda$ so that most of 
the power is associated with small wavenumbers $w< \Lambda$. Then, in the limit 
$k\gg\Lambda$, the evolution of a given mode $\vk$ is governed by the contributions of 
small wavenumbers, $k_1<\Lambda$ (whence $k_2 \simeq k$) or $k_2<\Lambda$ (whence 
$k_1 \simeq k$), in the right hand side of Eqs.(\ref{divkappaomega}) and 
(\ref{rotkappaomega}), that are further assumed to be in the linear regime.

The motion equations for the high-$\vk$ modes then form a set of 
{\sl linear} equations in presence of a random background described 
by the collection of the low-$\vw_{j}$ modes.
This leaves us with still a complicated system of equations to solve. A 
dramatic simplification can further be made because of the high-$k$ limit.
Indeed, since $k\gg w_j$, $\vw_{j}$ denoting the incoming linear wave modes,
the wave mode $\vk$ is almost left unchanged along the principal line 
(in other words one is entitled to replace $\Dirac(\vk'+\vw_j-\vk)$ in each vertex 
point by $\Dirac(\vk'-\vk)$). In this context, the motion equations that describe 
the mode evolution, Eqs.(\ref{divkappaomega}) and (\ref{rotkappaomega}), can by 
approximated by,
\begin{widetext}
\beqa
\kappa''(\vk,\eta)+\frac{1}{2}\kappa'(\vk,\eta)-\frac{3}{2}\kappa(\vk,\eta) & = &
\int \dd\vw\ \kappa_{L}(\vw,\eta)
\left\{ \alpha(\vk,\vw) \left[ \kappa''(\vk,\eta)+\frac{1}{2}\kappa'(\vk,\eta) \right] \right. 
\nonumber \\
&&  \left. + \beta(\vk,\vw) \left[\omega''(\vk,\eta)+\frac{1}{2}\omega'(\vk,\eta)\right] \right\} ,
\label{kappalin}\\
\omega'(\vk,\eta) &=& \int \dd\vw\ \kappa_{L}(\vw,\eta)
\{ \alpha(\vk,\vw) [\omega'(\vk,\eta)-\omega(\vk,\eta)] + \beta(\vk,\vw) [\kappa(\vk,\eta)-\kappa'(\vk,\eta)] \} ,
\label{omegalin}
\eeqa
so that high-$k$ modes now evolve {\sl independently} on one another.
One can easily check that the solution of Eqs.(\ref{kappalin})-(\ref{omegalin}), 
written as a perturbative series over $\kappa_{0}$, gives back the principal-path 
diagrams described above (here with the approximation $\vk'=\vk$).

Of course, we could apply the same procedure to the equations of motion 
(\ref{intdivkappaomega}),(\ref{introtkappaomega}), written in the integral form
\cite{2007A&A..476..31V,2007arXiv0711.3407V}.
This is equivalent to the differential form used above, but it is less convenient
for practical purposes. For instance, it is easier to solve numerically the
differential equations (\ref{kappalin})-(\ref{omegalin}) than their integral
equivalents which require the computation of an integral over all past values
to advance to the next time-step.

Then, we note from Eqs.(\ref{kappalin})-(\ref{omegalin}) that all contributions 
from the incoming waves $\kappa_L(\vw_j)$ can be factorized out and resummed in 
two distinct bundles of waves, $\ha$ and $\hb$, defined as,
\beq
\ha(\vk)= \int \dd\vw\ \kappa_{0}(\vw) \alpha(\vk,\vw) ,\ \ \ \ 
\hb(\vk)= \int \dd\vw\ \kappa_{0}(\vw) \beta(\vk,\vw),
\label{hbdef}
\eeq
such that Eqs.(\ref{kappalin})-(\ref{omegalin}) now read,
\beqa
\kappa''(\vk,\eta)+\frac{1}{2}\kappa'(\vk,\eta)-\frac{3}{2}\kappa(\vk,\eta) &=& 
e^{\eta} \ha(\vk) \left(\kappa''(\vk,\eta)+\frac{1}{2}\kappa'(\vk,\eta)\right) 
+ e^{\eta} \hb(\vk) \left(\omega''(\vk,\eta)+\frac{1}{2}\omega'(\vk,\eta)\right)
\label{kappahahb} \\
\omega'(\vk,\eta) &=& - e^{\eta} \hb(\vk) (\kappa'(\vk,\eta)-\kappa(\vk,\eta)) 
+ e^{\eta} \ha(\vk) (\omega'(\vk,\eta)-\omega(\vk,\eta)).
\label{omegahahb}
\eeqa
\end{widetext}

In other words, the fields $\kappa(\vk)$ and $\omega(\vk)$ depend on the 
linear modes only through the combinations $\ha$ and $\hb$. This introduces a dramatic 
simplification because then the ensemble average of Eq.(\ref{Gdef}) can be performed 
through a simple average over the two variables $\ha$ and $\hb$. 
Since Eqs.(\ref{kappahahb})-(\ref{omegahahb}) are linear the solution is proportional
to $\kappa_{0}(\vk)$. It is convenient to write it as,
\beqa
\kappa(\vk,\eta) = e^{\eta} \kappa_{0}(\vk)\ \hk(\eta) ,
\label{hkdef}\\
\omega(\vk,\eta) = e^{\eta} \kappa_{0}(\vk)\ \ho(\eta) .
\label{hodef}
\eeqa
As a consequence, we have
\beq
G(k,\eta)= e^{\eta} \hG(\eta) \;\;\; \mbox{with} \;\;\; \hG(\eta)= \lag \hk(\eta) \rag .
\label{hGdef}
\eeq
We can already note that because $\ha$ and $\hb$ depend on the direction of $\vk$ only,
$\hG(\eta)$ will be completely independent of $\vk$ (since a priori it could only 
depend on its norm).

In the last equation (\ref{hGdef}) the ensemble average now reduces to the computation 
of the expectation value of $\hk(\eta)$ with respect to the distribution of $\ha$ 
and $\hb$.
We then need to explore a bit more the statistical properties of $\ha$ and $\hb$.
Using the fact that the linear density field $\delta_{L}(\vq)=\kappa_{L}(\vq)$ 
is real, hence $\kappa_{0}(\vw)^*=\kappa_{0}(-\vw)$, it can be easily checked that 
$\ha$ and $\hb$ are real numbers. Moreover, we can see from 
Eqs.(\ref{hbdef}) that they are independent Gaussian random variables with:
\beq
\lag\ha^2\rag= 3\sigma_{2}^2 , \;\; \lag\hb^2\rag= \sigma_{2}^2, \;\; \lag\ha\hb\rag= 0,
\label{hab2}
\eeq
with,
\beq
\sigma_{2}^2= \frac{\pi}{4} \int_0^{\infty} \dd w\,  w P_{0}(w) .
\label{sigdef}
\eeq
Note that $8\sigma_2^2$ is also the variance of the density contrast 
$\lag\delta(\vx)^2\rag$.
The joint distribution function of $\ha$ and $\hb$ is then 
\begin{equation}
\mP(\ha,\hb)\,\dd\ha\,\dd\hb= \frac{\dd\ha\,\dd\hb}{\sqrt{3}2\pi\sigma_{2}^2}
\exp\left[{-\frac{\ha^2}{6\sigma_{2}^2}-\frac{\hb^2}{2\sigma_{2}^2}}\right] .
\label{hahbdistr}
\end{equation}

As a result we simply have,
\beq
\hG(\eta) = \int_{-\infty}^{\infty} \hk(\eta;\ha,\hb) \mP(\ha,\hb)\,\dd\ha\,\dd\hb,
\label{hGint}
\eeq
where $\hk(\eta;\ha,\hb)$ is the solution of the system 
(\ref{kappahahb})-(\ref{omegahahb}), written in terms of $\hk$ and $\ho$, 
parameterized by the coefficients $\ha,\hb$. 
The calculation of  the propagator can take advantage of the symmetries (\ref{parity}). 
In particular, we have:
\beq
\hk(\eta;\ha,\hb)=\hk(\eta;\ha,-\hb), \;\;  \ho(\eta;\ha,\hb)=-\ho(\eta;\ha,-\hb) .
\label{signbeta}
\eeq
We can also note that for $\mu>0$,
\beqa
\hk(\eta;\mu\ha,\mu\hb) &=&  \hk(\eta+\ln \mu;\ha,\hb) \label{scalehamu}\\
\ho(\eta;\mu\ha,\mu\hb) &=&  \ho(\eta+\ln \mu;\ha,\hb).\label{scalehbmu}
\eeqa

\subsubsection{Behavior of the propagator $G(k,\eta)$}
\label{Behavior}

The asymptotic behavior of $G(k,\eta)$ is intimately related to the behavior of the 
solutions of (\ref{kappahahb})-(\ref{omegahahb}) for finite values of the parameters 
$\ha$ and $\hb$. This can be inferred by inspection of these differential equations.
Thus, looking for an asymptotic power-law solution, $\hk\sim \hki e^{\nu\eta}$ and
$\ho\sim\hoi e^{\nu\eta}$, in the limit of large $\eta$
where the right hand side dominates in Eqs. (\ref{kappahahb})-(\ref{omegahahb}), 
we obtain the condition
\beqa
\lefteqn{\left| \bea{cc} \ha (\nu+1) (\nu+\frac{3}{2}) & \hb (\nu+1) (\nu+\frac{3}{2}) \\
-\hb \nu & \ha \nu \ea \right| } \nonumber \\
&& = (\ha^2+\hb^2) \nu (\nu+1) (\nu+\frac{3}{2}) = 0 ,
\label{Detnu}
\eeqa
which gives the asymptotic modes:
\beq
\nu_1=0 , \;\;\; \nu_2= -1, \;\;\; \nu_3= -\frac{3}{2} .
\label{nu+-}
\eeq
In fact, the mode $\nu_2$ can be removed since Eq.(\ref{kappahahb}) can be integrated 
once, as shown in Eq.(\ref{hkD1}) in the appendix.
Therefore, when $\eta\gg 1$, $\hk$ and $\ho$ are expected to be constant
(their value depending on the parameters $\ha$ and $\hb$ in a complicated way) 
because of the mode $\nu_1$.
This implies that the propagator $\hG$ obtained from the Gaussian integration 
(\ref{hGint}) must also be constant at late time. This expected behavior assumes 
that the differential equations (\ref{kappahahb})-(\ref{omegahahb}) do not encounter 
a singularity at a finite time $\eta$. We can check that 
Eqs. (\ref{kappahahb})-(\ref{omegahahb}) do not show explicit singularities associated 
with zeros of the coefficient of the higher-order terms. Indeed, the determinant of
the coefficients of highest-order derivatives reads as
\beq
\left| \bea{cc} 1-\ha e^{\eta} & -\hb e^{\eta} \\ 
\hb e^{\eta} & 1-\ha e^{\eta} \ea \right|  
= (1-\ha e^{\eta})^2 + (\hb e^{\eta})^2 
\label{DetD}
\eeq
which never vanishes if $\hb\neq 0$. We have checked numerically that
the system of differential equations (\ref{kappahahb})-(\ref{omegahahb}) obeys the 
behavior described above, with no singularity and a constant asymptote a late
time. This is depicted in Fig.~\ref{figkappa2D} with a 2D plot of 
$\hk(\eta;\ha,\hb)$ over the plane $(\ha,\hb)$ at time $\eta=0$
(this 2D plot is sufficient to fully determine the behavior of $\hk(\eta;\ha,\hb)$ 
thanks to the scaling law (\ref{scalehamu})).

\begin{figure}
\centerline{\epsfig {figure=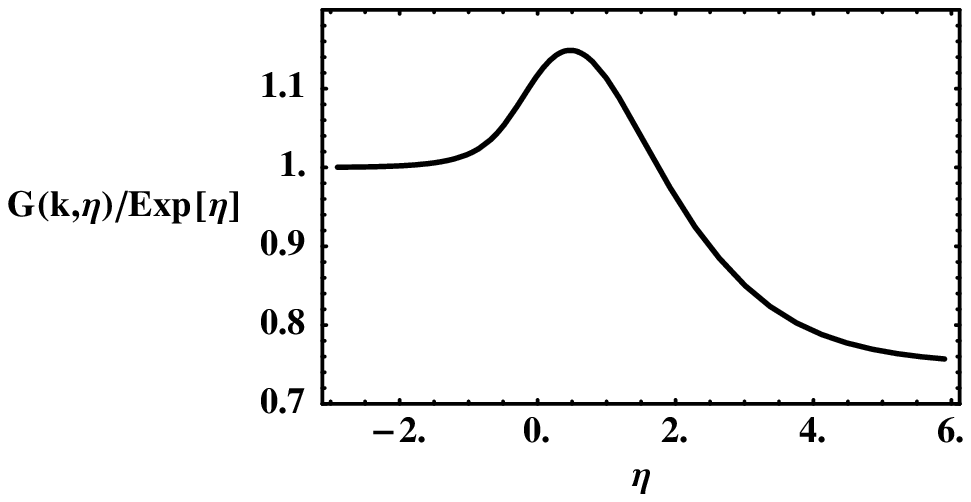,width=8cm}}
\caption{The propagator $\hG(\eta)$ as a function of $\eta$. It is obtained from
Eq.(\ref{hGint}) for a variance $\sigma_{2}=1$. Note that $\hG(\eta)$ depends only 
on the reduced variable $\eta+\ln \sigma_{2}$ as a consequence 
of (\ref{scalehamu}-\ref{scalehbmu}).} 
\label{figG2D}
\end{figure}

We show in Fig.~\ref{figG2D} our results for the propagator $\hG(\eta)$ obtained 
from the numerical integration of Eq.(\ref{hGint}). We can see that it first grows 
until it reaches a maximum at $\eta\sim 0$ and next decreases to converge to a constant 
$\hG(\eta=\infty) \sim 0.8$, a behavior qualitatively in agreement with the discussion 
above. Note that at early times the rise of $\hG$ means that the propagator 
$G(k,\eta)$ grows faster than the linear prediction (\ref{GL}), until $\eta \sim 1$.

The 2D case described above illustrates the power of the method based on associating
the series of principal-path diagrams with a linear dynamics as in 
Eqs.(\ref{kappahahb})-(\ref{omegahahb}). Indeed, in the high-$k$ limit the dependence
on initial conditions is reduced to a few random parameters (here $\ha$ and $\hb$),
as can also be read from the diagrams of Figs.~\ref{KappaExpansion}. Then, the
ensemble average is reduced to ordinary integrals (\ref{hGint}) (instead of path 
integrals over the field $\kappa_{0}(\vk)$) and the resummation associated with
the infinite series of diagrams is obtained by computing the exact solution of the
differential equations Eqs.(\ref{kappahahb})-(\ref{omegahahb}). Both steps can be
performed numerically, as above, since they only involve ordinary integrals and
differential equations (instead of functionals of fields). This allows us to compute
the propagator $G(k,\eta)$ even when the diagrammatic series cannot be exactly resummed
by analytical formulae (which corresponds to the case when the differential equations
(\ref{kappahahb})-(\ref{omegahahb}) have no known explicit solutions).
Moreover, even in this case, we can obtain exact analytical results for the late-time
non-perturbative behavior of the propagator, as in Eqs.(\ref{nu+-}), by direct
inspection of the effective linear equations of motion 
(\ref{kappahahb})-(\ref{omegahahb}).

We further discuss the properties of the system (\ref{kappahahb})-(\ref{omegahahb})
in appendix~\ref{Alternative-equations}. In particular, we show that taking into
account the vorticity (i.e. $\hb\neq 0$ and $\ho\neq 0$) is necessary to obtain
a well-behaved propagator at late times (otherwise a divergence appears), and that
the perturbative series over powers of $\kappa_{0}$ is probably only asymptotic 
(i.e. with zero radius of convergence).

\section{3D dynamics}
\label{3D-dynamics}

We now consider the case of the full 3D dynamics. This leads to slightly more
intricate expressions as we have a few more degrees of freedom but we can still
follow the analysis described in Sect.~\ref{2D-dynamics} for the simpler
2D dynamics. Moreover, we shall find that the results obtained
in Sect.~\ref{2D-dynamics} remain valid.

First, as in Eq.(\ref{chilambda}), we can decompose the displacement field
over a curl-free part $\chi$ and a divergence-less part $\vlambda$ as
\beq
\Psi = \left( \bea{c} \frac{\dr\chi}{\dr q_1} + \frac{\dr\lambda_3}{\dr q_2} 
- \frac{\dr\lambda_2}{\dr q_3}  \\
\frac{\dr\chi}{\dr q_2} + \frac{\dr\lambda_1}{\dr q_3} -
\frac{\dr\lambda_3}{\dr q_1} \\ 
\frac{\dr\chi}{\dr q_3} + \frac{\dr\lambda_2}{\dr q_1} -
\frac{\dr\lambda_1}{\dr q_2}  \ea \right) 
= \nabla_{\vq} . \chi + \nabla_{\vq} \times \vlambda  .
\label{chilambda3D}
\eeq
Thus, the rotational part $\vlambda$ has now two degrees of freedom: there are
three components $\lambda_1,\lambda_2,\lambda_3$, but the divergence of $\vlambda$
does not contribute and can be set to zero. As in 
Eqs.(\ref{kappachi})-(\ref{omegalambda}) we define the divergence $-\kappa$ and
the vorticity $\vomega$ by
\beq
\kappa= - \nabla_{\vq}^2 \chi , \;\;\; \vomega = - \nabla_{\vq}^2 \vlambda .
\label{omegalambda3D}
\eeq
Then, the spatial derivatives of the displacement field read in Fourier space as
\beq
\frac{\dr\Psi_i}{\dr q_j} = \Psi_{i,j}(\vk) = - \frac{k_ik_j}{k^2} \kappa(\vk) 
- \epsilon_{ilm} \frac{k_jk_l}{k^2} \omega_m(\vk) ,
\label{Psiij3D}
\eeq
where $\epsilon_{ilm}$ is the Levi-Civita symbol.
Then, the equation of motion (\ref{divmatrix}) reads in Fourier space as
\begin{widetext}
\beqa
\lefteqn{\hspace{-0.6cm} \kappa''+\frac{1}{2}\kappa'-\frac{3}{2}\kappa = 
\int \dd\vk_1\dd\vk_2 \Dirac(\vk_1+\vk_2-\vk)
\left\{ \frac{k_1^2k_2^2-(\vk_1.\vk_2)^2}{k_1^2k_2^2} \kappa_1 
(\kappa_2''+\frac{1}{2}\kappa_2'-\frac{3}{4}\kappa_2) 
- \frac{(\vk_1.\vk_2)}{k_1^2k_2^2} [\vk_2.(\vk_1\times\vomega_1)] 
(\kappa_2''+\frac{1}{2}\kappa_2'-\frac{3}{2}\kappa_2) \right. } \nonumber \\
&& \hspace{-1.2cm} \left. - \frac{(\vk_1.\vk_2)}{k_1^2k_2^2} \kappa_1 
[\vk_1.(\vk_2\times (\vomega_2''+\frac{1}{2}\vomega_2'))] \right\} 
- \int \dd\vk_1\dd\vk_2\dd\vk_3 \Dirac(\vk_1+\vk_2+\vk_3-\vk) \left\{
\frac{\det(\vk_1,\vk_2,\vk_3)^2}{2k_1^2k_2^2k_3^2} \kappa_1 \kappa_2 
(\kappa_3''+\frac{1}{2}\kappa_3'-\frac{1}{2}\kappa_3) \right. \nonumber \\
&& \hspace{-1.2cm} \left. + \frac{\det(\vk_1,\vk_2,\vk_3)}{k_1^2k_2^2k_3^2} 
[(\vk_2\times\vk_3).(\vk_1\times\vomega_1)] \kappa_2 
(\kappa_3''+\frac{1}{2}\kappa_3'-\frac{3}{4}\kappa_3)
+ \frac{\det(\vk_1,\vk_2,\vk_3)}{2k_1^2k_2^2k_3^2} \kappa_1 \kappa_2
[(\vk_1\times\vk_2).(\vk_3\times(\vomega_3''+\frac{1}{2}\vomega_3'))] 
\right\} \! + \! ..
\label{divkappaomega3D}
\eeqa
\end{widetext}
where the dots stand for terms of order $\omega^2$ and $\omega^3$.
We do not write these terms here since they will not contribute to the high-$k$
approximation. The determinant $\det(\vk_1,\vk_2,\vk_3)$ introduced in
Eq.(\ref{divkappaomega3D}) is the determinant of the $3\times 3$ matrix obtained
by putting the coordinates of the vectors $\vk_1,\vk_2$ and $\vk_3$, in the three
columns. It is also given by:
\beq
\det(\vk_1,\vk_2,\vk_3) = (\vk_1 \times \vk_2).\vk_3
\label{det3D}
\eeq
Note that Eq.(\ref{divkappaomega3D}) is now cubic over $\Psi$, hence over 
$\kappa,\omega$. For the constraints associated with the curl-free condition
(\ref{rotPsi}) we can use Eq.(\ref{rotmatrix2}) which is still quadratic.
This gives:
\begin{widetext}
\beq
\frac{(\vk\times\vomega')\times\vk}{k^2} = \int 
\dd\vk_1\dd\vk_2 \Dirac(\vk_1+\vk_2-\vk) \frac{\vk_1\times\vk_2}{k_1^2k_2^2}
\left\{ (\vk_1.\vk_2) \kappa_1 \kappa_2' + \kappa_1 
[\vk_1.(\vk_2\times\vomega_2')] + [\vk_2.(\vk_1\times\vomega_1)] \kappa_2' 
\right\} + ..
\label{rotkappaomega3D} 
\eeq
\end{widetext}
where the dots stand for terms of order $\omega^2$. We can check that only the
combination $\vk\times\vomega$ appears in 
Eqs.(\ref{divkappaomega3D})-(\ref{rotkappaomega3D}).
Moreover, as in the 2D case where Eqs.(\ref{divkappaomega})-(\ref{rotkappaomega}) 
obeyed the parity symmetry (\ref{parity}), we can check that 
Eqs.(\ref{divkappaomega3D})-(\ref{rotkappaomega3D}) are consistent with the
parity symmetry
\beq
\mP :\;\; \kappa \rightarrow \kappa, \;\;\; \vomega \rightarrow \vomega  .
\label{parity3D}
\eeq
In agreement with Eq.(\ref{chilambda3D}), $\vlambda$ and $\vomega$ are 
pseudovectors.

As in Sect.~\ref{Resummation}, the resummation associated with principal-path 
diagrams can be read from Eqs.(\ref{divkappaomega3D})-(\ref{rotkappaomega3D})
by linearizing over $\kappa,\vomega$. This yields
\begin{widetext}
\beqa
\lefteqn{\hspace{-0.3cm}\kappa''+\frac{1}{2}\kappa'-\frac{3}{2}\kappa = 
\int \dd\vw e^{\eta} \kappa_{0}(\vw) \left\{ \frac{k^2w^2-(\vk.\vw)^2}{k^2w^2} 
(\kappa''+\frac{1}{2}\kappa') - \frac{(\vk.\vw)}{k^2w^2} 
[\vw.(\vk\times(\vomega''+\frac{1}{2}\vomega'))] \right\} } \nonumber \\
&& \hspace{-0.6cm} -  \int \dd\vw\dd\vu e^{2\eta} \kappa_{0}(\vw) \kappa_{0}(\vu)
\left\{ \frac{\det(\vk,\vw,\vu)^2}{2k^2w^2u^2} 
(\kappa''+\frac{1}{2}\kappa'+\frac{3}{2}\kappa) 
+ \frac{\det(\vk,\vw,\vu)}{2k^2w^2u^2} 
[(\vw\times\vu).(\vk\times(\vomega''+\frac{1}{2}\vomega'+\frac{3}{2}\vomega))] 
\right\}
\label{kappalin3D}
\eeqa
\end{widetext}
and
\beqa
\lefteqn{ \frac{(\vk\times\vomega')\times\vk}{k^2} = \int \dd\vw e^{\eta} 
\kappa_{0}(\vw) \frac{\vw\times\vk}{k^2w^2} } \nonumber \\
&& \times \left\{ (\vk.\vw) (\kappa'-\kappa) + [\vw.(\vk\times(\vomega'-\vomega))] 
\right\} .
\label{omegalin3D} 
\eeqa
As for the 2D case, each mode $\kappa(\vk),\vomega(\vk)$ evolves independently
of other high-$k$ modes and the dependence on the initial field $\kappa_{0}$
is reduced to a few random parameters that can be written as integrals over
$\kappa_{0}$. In order to make further progress, it is convenient to write
Eqs.(\ref{kappalin3D})-(\ref{omegalin3D}) in terms of coordinates. Without any loss
of generality, we can choose $\vk$ along the axis $\ve_1$, and $\vomega$ in the
plane $(\ve_2,\ve_3)$. Then, Eqs.(\ref{kappalin3D})-(\ref{omegalin3D}) read as
\begin{widetext}
\beqa
\kappa''+\frac{1}{2}\kappa'-\frac{3}{2}\kappa & = & e^{\eta} (\tau_{22}+\tau_{33}) 
(\kappa''+\frac{1}{2}\kappa') - e^{\eta} \tau_{13} (\omega_2''+\frac{1}{2}\omega_2') 
+ e^{\eta} \tau_{12} (\omega_3''+\frac{1}{2}\omega_3') - e^{2\eta} 
(\tau_{22}\tau_{33}-\tau_{23}^2) (\kappa''+\frac{1}{2}\kappa'+\frac{3}{2}\kappa)
\nonumber \\
&& - e^{2\eta} (\tau_{12}\tau_{23}-\tau_{22}\tau_{13}) 
(\omega_2''+\frac{1}{2}\omega_2'+\frac{3}{2}\omega_2) 
+  e^{2\eta} (\tau_{13}\tau_{23}-\tau_{33}\tau_{12}) 
(\omega_3''+\frac{1}{2}\omega_3'+\frac{3}{2}\omega_3) ,
\label{kappaAij}
\eeqa
\end{widetext}
and:
\beq
\omega_2'= e^{\eta} \tau_{13} (\kappa'-\kappa) + e^{\eta} \tau_{33} (\omega_2'-\omega_2)
- e^{\eta} \tau_{23} (\omega_3'-\omega_3) ,
\label{omega2Aij}
\eeq
\beq
\omega_3'= -e^{\eta} \tau_{12} (\kappa'-\kappa) - e^{\eta} \tau_{23} (\omega_2'-\omega_2)
+ e^{\eta} \tau_{22} (\omega_3'-\omega_3) .
\label{omega3Aij}
\eeq
Here we introduced the symmetric parameters $\tau_{ij}$ defined by:
\beq
\tau_{ij} = \int\dd\vw\ \kappa_{0}(\vw) \frac{w_i\,w_j}{w^2} .
\label{Adef}
\eeq
Using the property $\kappa_{0}(\vw)^*=\kappa_{0}(-\vw)$, we can see that
the coefficients $\tau_{ij}$ are real random numbers.
We recover the two-dimensional case (\ref{kappahahb})-(\ref{omegahahb}) for
\beq
\tau_{i3}=0 , \;\; \omega_2=0, \;\; \ha= \tau_{22}, \;\; \hb= \tau_{12} ,
\eeq
or
\beq
\tau_{i2}=0 , \;\; \omega_3=0, \;\; \ha= \tau_{33}, \;\; \hb= -\tau_{13} .
\eeq
Note that there are several symmetry properties. Two symmetries extend the property 
(\ref{signbeta}) obtained for the 2D case. They read as
\beq
\tau_{13}\rightarrow -\tau_{13}, \; \tau_{23}\rightarrow -\tau_{23}, 
\; \omega_2 \rightarrow -\omega_2 ,
\label{signom2}
\eeq
and 
\beq
\tau_{12}\rightarrow -\tau_{12}, \; \tau_{23}\rightarrow -\tau_{23}, 
\; \omega_3 \rightarrow -\omega_3 ,
\label{signom3}
\eeq
where we only write the quantities that change under these two symmetries.
A further symmetry comes from the invariance over a coordinate rotation 
in the $(\ve_{2},\ve_{3})$ plane. 
To express it, we can define the following quantities,
\beqa
\tau&=&\frac{\tau_{22}+\tau_{33}}{2} , \\
\vv=v\ e^{\ii \theta_{v}}&=&\tau_{12}+\ii \tau_{13} , \\
\vgamma=\gamma\ e^{2\ii \theta_{\gamma}}&=&\frac{\tau_{22}-\tau_{33}}{2}+\ii \tau_{23} ,
\eeqa
which behave respectively like spin 0, 1 and 2 complex numbers with respect to
coordinate rotations in the $(\ve_{2},\ve_{3})$ plane. 
The ensemble average of those quantities can be expressed in terms of 
$\sigma_{3}^2$ defined as,
\begin{equation}
\sigma_{3}^2=\frac{8\pi}{15}\int \dd w \, w^2 P_{0}(w),
\label{sigma3D}
\end{equation}
with
\begin{equation}
\lag\tau^2\rag= \lag|\vv|^2\rag= \lag|\vgamma|^2\rag= \sigma_{3}^2,
\label{moments3D}
\end{equation}
while cross-correlations between these quantities vanish.
We also define the complex vorticity $\omega$ as
\beq
\omega = - \omega_3+\ii \omega_2 ,
\label{omegaspin}
\eeq
which is of spin 1 like $\vv$. Here we use the fact that, as in the 2D case,
Eqs.(\ref{kappaAij})-(\ref{omega3Aij}) are linear so that we can factorize
a factor $\kappa_0(\vk)$, as in Eqs.(\ref{hkdef})-(\ref{hodef}). Then,
the reduced quantities $\kappa/\kappa_0,\omega_i/\kappa_0$, are real (since
the coefficients $\tau_{ij}$ of Eq.(\ref{Adef}) are real)
so that the complex vorticity (\ref{omegaspin}) fully determines the doublet
$\{\omega_2,\omega_3\}$. Then, the two equations 
(\ref{omega2Aij})-(\ref{omega3Aij}) can be gathered into
\beq
\omega'= e^{\eta} \vv (\kappa'-\kappa) + e^{\eta} \tau (\omega'-\omega)
+ e^{\eta} \vgamma (\omega'-\omega)^* ,
\label{omegatij}
\eeq
whereas Eq.(\ref{kappaAij}) reads as
\begin{widetext}
\beqa
\kappa''+\frac{1}{2}\kappa'-\frac{3}{2}\kappa & = & e^{\eta} 2\tau 
(\kappa''+\frac{1}{2}\kappa') - e^{\eta} \frac{\vv^*}{2} (\omega''+\frac{1}{2}\omega') 
- e^{\eta} \frac{\vv}{2} (\omega''+\frac{1}{2}\omega')^* - e^{2\eta} 
(\tau^2-\vgamma\vgamma^*) (\kappa''+\frac{1}{2}\kappa'+\frac{3}{2}\kappa)
\nonumber \\
&& - e^{2\eta} \frac{\vv\vgamma^*-\tau\vv^*}{2} 
(\omega''+\frac{1}{2}\omega'+\frac{3}{2}\omega) 
- e^{2\eta} \frac{\vv^*\vgamma-\tau\vv}{2} 
(\omega''+\frac{1}{2}\omega'+\frac{3}{2}\omega)^* .
\label{kappatij}
\eeqa
\end{widetext}
We can see that all terms in Eq.(\ref{omegatij}) are of spin 1, whereas all terms
in Eq.(\ref{kappatij}) are of spin 0. This clearly shows that these equations are 
invariant through rotations in the $(\ve_{2},\ve_{3})$ plane. Moreover, we can check 
that both sides in Eq.(\ref{kappatij}) are real. Obviously, the results depend
only on the angle difference $\theta_v-\theta_{\gamma}$. 

Finally, the scaling laws (\ref{scalehamu})-(\ref{scalehbmu}) also extend 
to the 3D case as
\beq
\mu > 0 : \;\;\; \tau_{ij} \rightarrow \mu \tau_{ij} , \;\; 
\eta \rightarrow \eta - \ln\mu .
\label{scaling3D}
\eeq

\begin{figure}
\centerline{\epsfig {figure=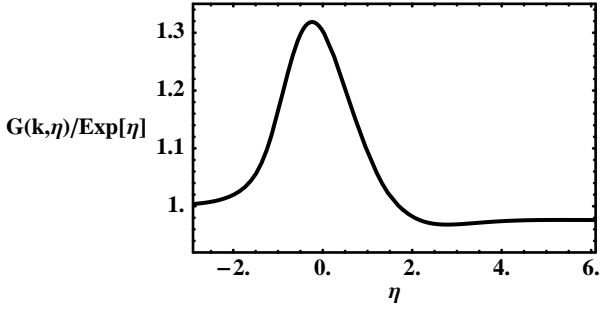,width=8cm}}
\caption{The propagator $\hG(\eta)$ as a function of $\eta$ for the 3D dynamics. 
It is obtained for a variance $\sigma_{3}=1$.} 
\label{figG3D}
\end{figure}

As for the 2D case analized in Sect.~\ref{Behavior}, we can look for singularities
associated with zeros of the determinant of the coefficients of higher-order derivatives.
This gives from Eqs.(\ref{omegatij})-(\ref{kappatij}):
\beq
\Delta = \left[ (1-e^{\eta}\tau)^2-e^{2\eta}|\vgamma|^2\right]^2 
+ \left|e^{\eta}\vv+e^{2\eta} (\vv^*\vgamma-\tau\vv)\right|^2
\label{sing3D}
\eeq
The determinant $\Delta$ can only vanish if $|\vgamma|=0$, or $|\vv|=0$, or
$\theta_v-\theta_{\gamma}=n \pi/2$ with $n$ integer, which is a region of
zero measure in the space spanned by the coefficients $\tau,\vv$, and $\vgamma$.
We have checked numerically that the differential system is otherwise 
well-behaved and the ensemble averages lead to well-defined quantities. 
As in Eqs.(\ref{Detnu})-(\ref{nu+-}), the asymptotic behavior of the solutions
$\kappa,\omega_2,\omega_3$, can be read from the differential equations 
(\ref{kappaAij})-(\ref{omega3Aij}) by looking for asymptotic power-laws.
This yields for the reduced variables $\hk,\ho_2,\ho_3$, defined as in 
Eqs.(\ref{hkdef})-(\ref{hodef}), the three asymptotic modes:
\beq
\nu_1=0, \;\; \nu_2=\frac{-5-\ii\sqrt{23}}{4} , \;\; \nu_3=\frac{-5+\ii\sqrt{23}}{4} .
\label{nu3D}
\eeq
Therefore, the reduced propagator $\hG(\eta)$ must go to a constant at late times,
as for the 2D case. Our numerical results are shown in Fig.~\ref{figG3D}
and we can see that they agree with this analysis.
Thus, it appears that the 3D propagator exhibits the same features as the 2D case,
with an early rise that is faster than the linear prediction and a late-time
behavior that follows the linear power-law $G(\eta) \sim e^{\eta}$ 
(with a ``renormalized'' amplitude that is smaller than unity).

As for the 2D case (see Eq.(\ref{sigdef})), the key quantity $\sigma_{3}^2$ that
measures the amplitude of the fluctuations and the state of gravitational 
clustering, see Eqs.(\ref{sigma3D})-(\ref{moments3D}), is proportional to the 
variance of the density field $\lag\delta(\vx)^2\rag$. We can note that for
a CDM power spectrum it shows a logarithmic UV divergence (since 
$P_{0}(w) \sim w^{-3}$ at high $w$). Therefore, our results rigorously apply
to linear power spectra with a high-$k$ cutoff such that $\sigma_{3}^2$ is
finite. However, since the fluid description does not hold beyond shell-crossing
it could be argued that integrals such as (\ref{sigma3D}) should be cut at the scale
associated with the transition to nonlinearity in any case.
On the other hand, within the high-$k$ approximation studied in this article,
the quantity $\sigma_3$ of Eq.(\ref{sigma3D}) should be interpreted as the
variance of the larger-scale density contrast, rather than the variance of
the one-point density contrast. Indeed, we can see from Eq.(\ref{Adef}) that
the quantity which governs the coefficients $\tau_{ij}$ is the density contrast
at the origin $\delta(\vx=0)$ (discarding the angular dependence associated
with $w_iw_j/w^2$). This in turn gives rise to Eq.(\ref{sigma3D}). Mathematically,
the specific role played by the origin is related to the breakdown of the invariance
through translations entailed by the approximation $\delta_D(\vk_1+\vk_2-\vk) \simeq
\delta_D(\vk_1-\vk)$ discussed in section~\ref{Resummation}. However, it is clear
that within this approximation, based on a separation of scales between low 
wavenumbers $w<\Lambda$ and high wavenumbers $k \gg \Lambda$, any point located
at a distance below $1/\Lambda$ from the origin could as well be chosen as a 
reference. In other words, within this high-$k$ approximation, $\sigma_{3}^2$
should be understood as the variance of the larger-scale density contrast,
associated with wavenumbers $w<\Lambda$ (and $\Lambda <k$). 
Then, in Eq.(\ref{sigma3D}) 
we relaxed the cutoff $\Lambda$, which is valid for linear power spectra with
small high-$k$ power so that the integral converges (and the high-$k$ approximation
discussed in section~\ref{Resummation} can make sense). We can see that
CDM power spectra are at the limit of applicability of this approximation.
  
We can note that the same features apply to the Eulerian description, except
that instead of the larger-scale density contrast the key quantity is the
larger-scale velocity. Then, it happens that CDM power spectra are fully within
the range where the velocity integral analogous to Eq.(\ref{sigma3D}) converges.

\section{Discussions}
\label{Discussions}

We have applied to the Lagrangian formalism a resummation scheme developed in
\cite{2006PhRvD..73f3520C} within the Eulerian description. This is based on the
resummation of a certain type of diagrams, called ``principal-path diagrams'' in
\cite{2006PhRvD..73f3520C}, that may be expected to dominate the dynamics in
a high-$k$ limit. In the Eulerian case, these diagrams can be explicitly computed, 
order by order, and resummed, as one can recognize the exponential function in
the series expansion obtained in this manner. This leads to a Gaussian decay
of the form $e^{-e^{2\eta}k^2\sigma_v^2/2}$ at high $k$.

It is more difficult to apply the same method to the Lagrangian formalism, as the
diagrams have a slightly more intricate expression and one cannot identify from the
series a well-known mathematical function. However, as shown in 
\cite{2007A&A..476..31V}, it is possible to identify this resummation with the
solution of an effective linear dynamics. Then, instead of computing explicitly
all diagrams and next resumming their contributions (which amounts to solve
for this effective equation of motion as a perturbative series), one can directly
solve for this simpler dynamics. In this article, we have applied this technique
to the Lagrangian description. We have shown that it is very powerful as
it can be used even when no explicit analytical solutions can be found (but one
can still solve numerically the relevant differential equations). Moreover, 
even in such cases, it is possible to obtain the exact exponents (as defined by this 
partial resummation) of the late-time regime, by looking for the asymptotic modes
of the linear differential equations. Then, we have found a late-time
power-law behavior for the propagator, which actually simply follows the linear
growth $e^{\eta}$ albeit with a ``renormalized'' amplitude slightly smaller than
unity. This is quite different from the Gaussian decay obtained in the Eulerian
case.

For comparison, let us briefly recall how this method applies to the Eulerian
case \cite{2007A&A..476..31V}. In this case, the solution to the effective
linear dynamics can be derived explicitly and it reads as
\beq
\delta(\vk,\eta) = e^{\eta} \delta_{0}(\vk) e^{e^\eta \ha_E(\vk)} ,
\label{deltaE}
\eeq
with
\begin{equation}
\ha_E(\vk)=\int\dd^n\vw \frac{\vk.\vw}{w^2}\delta_{0}(\vw),
\label{haEDef}
\end{equation}
(using notations that straightforwardly extend those used throughout the paper). 
For Gaussian initial conditions the ensemble average of this expression can be easily 
computed. It leads to the following propagator,
\beq
G_E(k,\eta) = e^{\eta} \, e^{e^{2\eta} \lag\ha_E(\vk)^2\rag/2} .
\label{GE}
\eeq
Because of the $\vk$-dependence of $\ha_E(\vk)$, one obtains a Gaussian damping 
of the form $e^{-e^{2\eta} k^2 \sigma_v^2/2}$ at high $k$, with 
$\sigma_v^2=1/n\int(\dd^n\vw/w^2) P_{0}(w)$ ($n$ is here the number of space 
dimensions). As discussed above and shown in details in previous sections,
our calculations in Lagrangian space do not give a closed form for the 
propagator but allow nonetheless to describe its properties exhaustively.

Eulerian and Lagrangian calculations prove to lead to quantitatively very different 
results.
Whereas the decay found for the Eulerian case exhibits a Gaussian tail with a strong 
$k$-dependence, in Lagrangian variables the propagators are essentially $k$-independent
with no significant decay at late time. After a stage of accelerated growth, followed
by a transitory slow-down, the high-$k$ modes growth is indeed found to be simply 
slightly retarded and still growing as $e^{\eta}$ as the linear growth rate. The 
situation is the same in 2D and 3D cases. The delay is only slightly less important 
for the 3D case.
The independence on wavenumber $k$ in the Lagrangian case directly follows from
the fact that the kernels $\alpha$ and $\beta$ of Eqs.(\ref{defalpha})-(\ref{defbeta}), 
that appear in the 2D equations of motion (\ref{divkappaomega}) 
and (\ref{rotkappaomega}),
are homogeneous functions of their two arguments $\vk_1$ and $\vk_2$: they only depend
on relative angles. This also holds for the 3D dynamics, as can be checked in
Eqs.(\ref{divkappaomega3D})-(\ref{rotkappaomega3D}). Therefore, this property is not
restricted to the partial resummation associated with ``principal-path'' diagrams.
In a similar fashion, the
dependence on $k$ obtained in the Eulerian case is due to the non-homogeneous 
character of the kernels $\alpha$ and $\beta$ that appear in this framework, which can
also be seen in Eq.(\ref{haEDef}).

As seen in the previous sections, another distinctive feature of the Lagrangian 
description is the important role played by parity symmetries. Indeed, whereas
in the Eulerian framework the two quantities of interest, the density and the
velocity divergence, are true scalars, in the Lagrangian framework we must take into
account both curl-free and rotational parts of the displacement field (in
Lagrangian space $\vq$), as a curl-free Eulerian velocity field does not translate
into a Lagrangian curl-free displacement field beyond second order.
We have shown that keeping track of the vorticity degrees of freedom is necessary
to obtain a well-defined propagator in the nonlinear regime.

How to reconcile these results? Although Eulerian and Lagrangian descriptions are 
ultimately equivalent, the objects we have computed are clearly distinct. In the 
nonlinear regime modes in Lagrangian space cannot be directly mapped to those in 
Eulerian space. One should then not be too surprised to find quantitatively different 
results.
What we have computed here is in essence the leading effect of a random background 
on the growth of structures, assuming scales can be well separated (e.g. that the 
wavelength of the background modes are much larger than the modes of 
interest).
It turns out that the Eulerian modes are sensitive to the large-scale displacement 
field at leading order, whereas the Lagrangian modes are not. These large-scale displacement fields 
are responsible for the decay of the Eulerian correlators at large time separations. 
Indeed, the modes behave as if they were randomly advected by the large-scale
displacements \cite{2006PhRvD..73f3520C, 2007A&A..476..31V}. Basically, everything happens as if
small-scale structures
were moved around; and because they occupy a different location in real space, their correlation 
with the initial field decay. In Lagrangian space, modes are not affected by such 
displacements (by construction, the convergence $\kappa$ and the vorticity
$\omega$ are not sensitive to a uniform translation, being related to derivatives
of the displacement field taken as a function of the initial conditions). They are
more directly sensitive to the density field. Thus, as discussed in 
section~\ref{3D-dynamics}, whereas the Eulerian propagator is governed by the
amplitude of the larger-scale velocity, the Lagrangian propagator is governed by the
amplitude of the larger-scale density.
Then, the leading effect
resembles more a tidal effect.  What we have found is that modes are not disrupted by the
accumulation of those  tidal effects, at this order of the calculation, e.g., the
results displayed in Figs.~\ref{figG2D} and \ref{figG3D} suggest that there is no 
true loss of memory nor efficient relaxation associated with the gravitational dynamics.
It is not clear then how this loss of memory - which is expected to happen eventually 
in the nonlinear regime - could take place. Whether it can be described with the help of 
additional diagrams\footnote{We can note that including other diagrams would mean 
that the incoming waves are not necessarily in the linear regime. They would then 
have not only different statistical properties but also different time dependences.},
from terms beyond the high-$k$ limit; or whether we have to go beyond 
shell-crossing (which breaks the analyticity of the Jacobian) to capture possible relaxation effects,
is yet unclear.

\begin{acknowledgments}
It is a pleasure to thank Rom\'an Scoccimarro and Mart\'{\i}n Crocce for fruitful discussions and
suggestions. This work was supported in part by the French Programme National de Cosmology and by 
the French Agence National de la Recherche under grant BLAN07-1-212615.
\end{acknowledgments}

\appendix
\section{Alternative equations for the 2D dynamics}
\label{Alternative-equations}

Here we explore in a more details the properties of system 
(\ref{kappahahb})-(\ref{omegahahb}). With the change of variable $D=e^{\eta}$ 
it yields,
\beqa
D^2 \frac{\dd^2\hk}{\dd D^2} 
+ \frac{7}{2} D \frac{\dd\hk}{\dd D} 
&=& \ha D \left[ D^2 \frac{\dd^2\hk}{\dd D^2} + \frac{7}{2} D \frac{\dd\hk}{\dd D} 
+ \frac{3}{2} \hk \right] \nonumber \\
&& \hspace{-1cm}+ \hb D \left[ D^2 \frac{\dd^2\ho}{\dd D^2} 
+ \frac{7}{2} D \frac{\dd\ho}{\dd D} + \frac{3}{2} \ho \right]
\label{hkD}\\
D \frac{\dd\ho}{\dd D} + \ho& = & -\hb D^2 \frac{\dd\hk}{\dd D} + 
\ha D^2 \frac{\dd\ho}{\dd D}
\label{hoD}
\eeqa
with the initial conditions:
\beq
D\rightarrow 0 : \;\;\;\; \hk=1+\frac{3}{7} \ha D, \;\;\;\; \ho=0.
\label{hkho0}
\eeq
Equation (\ref{hkD}) can be integrated once to give:
\beq
D \frac{\dd\hk}{\dd D} + \frac{5}{2} \hk - \frac{5}{2} =
\ha D \left[ D \frac{\dd\hk}{\dd D} + \frac{3}{2} \hk \right] 
+ \hb D \left[ D \frac{\dd\ho}{\dd D} 
+ \frac{3}{2} \ho \right]
\label{hkD1}
\eeq
Then, eliminating $\ho$ from Eqs.(\ref{hoD})-(\ref{hkD1}) gives the second-order
equation for $\hk$:
\beqa
\lefteqn{ 2(1-3\ha D) D^2\frac{\dd^2\hk}{\dd D^2} + (7-15\ha D) D \frac{\dd\hk}{\dd D} }
\nonumber \\
&& + \frac{\ha D(1-\ha D)}{(1-\ha D)^2+(\hb D)^2} [ 12 \hk - 15 ] = 0
\label{hkD2}
\eeqa

\begin{figure*}
\centerline{
\begin{tabular}{lcr}
\epsfig{figure=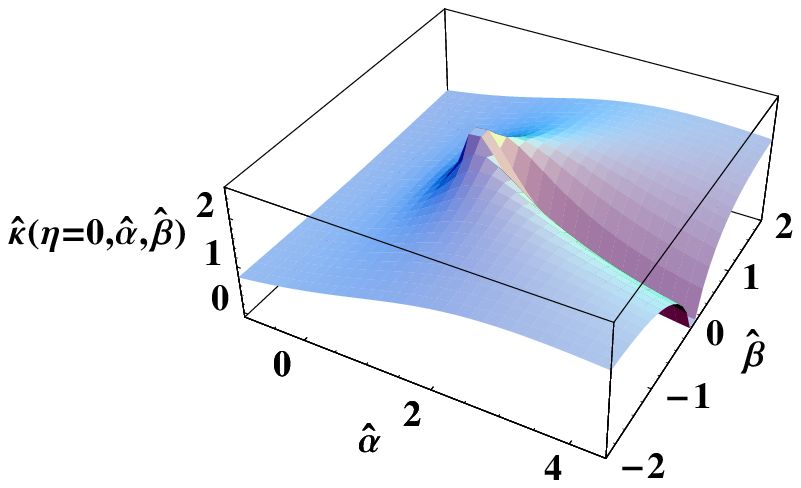,width=8cm}&\hspace{1.cm}&
\epsfig{figure=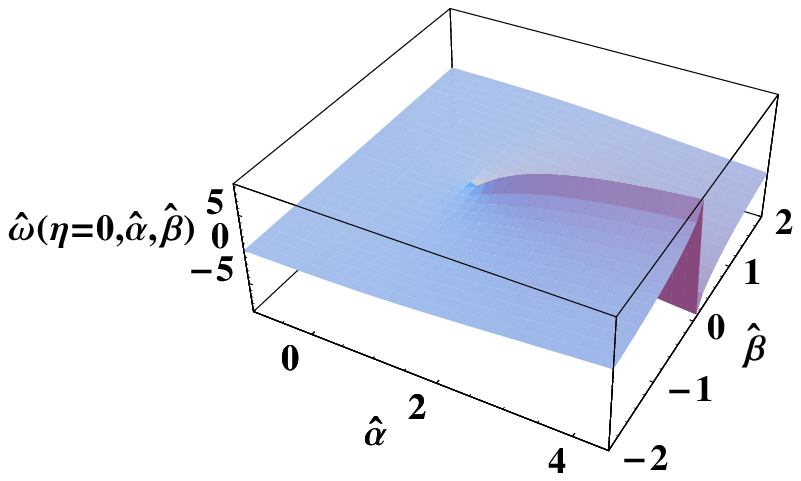,width=8cm}
\end{tabular}
}
\caption{The divergence $\hk(\eta=0;\ha,\hb)$ (left panel) and the vorticity 
(right panel) as a function of $\ha,\hb$. The divergence reaches a constant at 
large radius, $\sqrt{\ha^2+\hb^2}\to\infty$, for a fixed polar angle. 
The divergence is found to be continuous and even with respect to $\hb$; 
the vorticity is found to be discontinuous along the critical half-line
$e^{\eta}\ha>1$, $\hb=0$, and odd with respect to $\hb$.}
\label{figkappa2D}
\end{figure*}

We can note that for $\hb=0$ the divergent part $\hk$
decouples from the vorticity $\ho$ and the solution of Eq.(\ref{hkD}) can be written
as
\beq
\hk(D;\ha,0) = \,  _2\!F_1(1,3/2;7/2;\ha D) .
\label{hkha}
\eeq
It exhibits a singularity at the point $D=1/\ha$ (for $\ha>0$), in agreement with 
Eq.(\ref{DetD}), but it actually remains finite at this point and has a well-behaved 
analytic continuation beyond, as seen in Fig.~\ref{figkappa2D}.

Here we can note that writing the high-$k$ resummation in terms of the differential
equations (\ref{kappalin})-(\ref{omegalin}), and the propagator with the integral
representation (\ref{hGint}), is a key ingredient to obtain the asymptotic behaviors.
Indeed, computing $G(k,\eta)$ from its diagrammatic expansion, which amounts to
expand the integrand in Eq.(\ref{hGint}) over powers of $\ha$ and $\hb$, leads
to an asymptotic series with zero radius of convergence. For instance, for $\hb=0$
we directly obtain from Eq.(\ref{hkha})
\beq
\hk(D;\ha,0) = \sum_{p=0}^{\infty} \frac{15}{(2p+3)(2p+5)} \ha^p D^p ,
\label{hkseries}
\eeq
which gives after we average over $\ha$:
\beq
\lag\hk(D;\ha,0)\rag_{\ha} = \sum_{p=0}^{\infty} \frac{15 \; (2p-1)!!}{(4p+3)(4p+5)} 
(3\sigma_2^2)^{2p} D^{2p} .
\label{Gseries}
\eeq
This asymptotic series describes the early rise of $\hG$ but it cannot
give (without ambiguities) the late-time relaxation to a constant.

This behavior emerges because of the existence of the second degree of freedom
associated with the vorticity, $\hb$. To take it into account, one may look for 
a solution of Eqs.(\ref{hoD})-(\ref{hkD1}) as a perturbative
series over powers of $D$, as in Eq.(\ref{hkseries}). Then, computing the first
few terms or looking at simplified cases suggests that the nonzero variance
of $\hb$ decreases somewhat the coefficients of Eq.(\ref{Gseries}) but they
remain positive and fastly growing (it typically modifies Eq.(\ref{Gseries})
by changing the factor $(3\sigma_2^2)^{2p}$ into $(2\sigma_2^2)^{2p}$, because of
Eq.(\ref{hab2})). Thus, as expected the vorticity slows down the rise of the
propagator $G(k,\eta)$ but its magnitude is not sufficient to make it decay with
respect to the linear propagator at early times. Nevertheless, it is necessary
to take into account the vorticity to obtain the late-time behavior of $G(k,\eta)$.

\bibliography{GrandesStructures}

\begin{thebibliography}{20}
\expandafter\ifx\csname natexlab\endcsname\relax\def\natexlab#1{#1}\fi
\expandafter\ifx\csname bibnamefont\endcsname\relax
  \def\bibnamefont#1{#1}\fi
\expandafter\ifx\csname bibfnamefont\endcsname\relax
  \def\bibfnamefont#1{#1}\fi
\expandafter\ifx\csname citenamefont\endcsname\relax
  \def\citenamefont#1{#1}\fi
\expandafter\ifx\csname url\endcsname\relax
  \def\url#1{\texttt{#1}}\fi
\expandafter\ifx\csname urlprefix\endcsname\relax\def\urlprefix{URL }\fi
\providecommand{\bibinfo}[2]{#2}
\providecommand{\eprint}[2][]{\url{#2}}

\bibitem[{\citenamefont{{Peebles}}(1980)}]{1980lssu.book.....P}
\bibinfo{author}{\bibfnamefont{P.~J.~E.} \bibnamefont{{Peebles}}},
  \emph{\bibinfo{title}{{The large-scale structure of the universe}}}
  (\bibinfo{publisher}{Research supported by the National Science
  Foundation.~Princeton, N.J., Princeton University Press, 1980.~435 p.},
  \bibinfo{year}{1980}).

\bibitem[{\citenamefont{{Bernardeau} et~al.}(2002)\citenamefont{{Bernardeau},
  {Colombi}, {Gazta{\~n}aga}, and {Scoccimarro}}}]{2002PhR...367....1B}
\bibinfo{author}{\bibfnamefont{F.}~\bibnamefont{{Bernardeau}}},
  \bibinfo{author}{\bibfnamefont{S.}~\bibnamefont{{Colombi}}},
  \bibinfo{author}{\bibfnamefont{E.}~\bibnamefont{{Gazta{\~n}aga}}},
  \bibnamefont{and}
  \bibinfo{author}{\bibfnamefont{R.}~\bibnamefont{{Scoccimarro}}},
  \bibinfo{journal}{\physrep} \textbf{\bibinfo{volume}{367}},
  \bibinfo{pages}{1} (\bibinfo{year}{2002}).

\bibitem[{\citenamefont{{Zel'Dovich}}(1970)}]{1970A&A.....5...84Z}
\bibinfo{author}{\bibfnamefont{Y.~B.} \bibnamefont{{Zel'Dovich}}},
  \bibinfo{journal}{\aap} \textbf{\bibinfo{volume}{5}}, \bibinfo{pages}{84}
  (\bibinfo{year}{1970}).

\bibitem[{\citenamefont{{Peacock} and {Dodds}}(1996)}]{1996MNRAS.280L..19P}
\bibinfo{author}{\bibfnamefont{J.~A.} \bibnamefont{{Peacock}}}
  \bibnamefont{and} \bibinfo{author}{\bibfnamefont{S.~J.}
  \bibnamefont{{Dodds}}}, \bibinfo{journal}{\mnras}
  \textbf{\bibinfo{volume}{280}}, \bibinfo{pages}{L19} (\bibinfo{year}{1996}).

\bibitem[{\citenamefont{{Smith} et~al.}(2003)\citenamefont{{Smith}, {Peacock},
  {Jenkins}, {White}, {Frenk}, {Pearce}, {Thomas}, {Efstathiou}, and
  {Couchman}}}]{2003MNRAS.341.1311S}
\bibinfo{author}{\bibfnamefont{R.~E.} \bibnamefont{{Smith}}},
  \bibinfo{author}{\bibfnamefont{J.~A.} \bibnamefont{{Peacock}}},
  \bibinfo{author}{\bibfnamefont{A.}~\bibnamefont{{Jenkins}}},
  \bibinfo{author}{\bibfnamefont{S.~D.~M.} \bibnamefont{{White}}},
  \bibinfo{author}{\bibfnamefont{C.~S.} \bibnamefont{{Frenk}}},
  \bibinfo{author}{\bibfnamefont{F.~R.} \bibnamefont{{Pearce}}},
  \bibinfo{author}{\bibfnamefont{P.~A.} \bibnamefont{{Thomas}}},
  \bibinfo{author}{\bibfnamefont{G.}~\bibnamefont{{Efstathiou}}},
  \bibnamefont{and} \bibinfo{author}{\bibfnamefont{H.~M.~P.}
  \bibnamefont{{Couchman}}}, \bibinfo{journal}{\mnras}
  \textbf{\bibinfo{volume}{341}}, \bibinfo{pages}{1311} (\bibinfo{year}{2003}),
  \eprint{arXiv:astro-ph/0207664}.

\bibitem[{\citenamefont{{Hamilton} et~al.}(1991)\citenamefont{{Hamilton},
  {Kumar}, {Lu}, and {Matthews}}}]{1991ApJ...374L...1H}
\bibinfo{author}{\bibfnamefont{A.~J.~S.} \bibnamefont{{Hamilton}}},
  \bibinfo{author}{\bibfnamefont{P.}~\bibnamefont{{Kumar}}},
  \bibinfo{author}{\bibfnamefont{E.}~\bibnamefont{{Lu}}}, \bibnamefont{and}
  \bibinfo{author}{\bibfnamefont{A.}~\bibnamefont{{Matthews}}},
  \bibinfo{journal}{\apjl} \textbf{\bibinfo{volume}{374}}, \bibinfo{pages}{L1}
  (\bibinfo{year}{1991}).

\bibitem[{\citenamefont{{Cooray} and {Sheth}}(2002)}]{2002PhR...372....1C}
\bibinfo{author}{\bibfnamefont{A.}~\bibnamefont{{Cooray}}} \bibnamefont{and}
  \bibinfo{author}{\bibfnamefont{R.}~\bibnamefont{{Sheth}}},
  \bibinfo{journal}{\physrep} \textbf{\bibinfo{volume}{372}},
  \bibinfo{pages}{1} (\bibinfo{year}{2002}), \eprint{astro-ph/0206508}.

\bibitem[{\citenamefont{{Crocce} and
  {Scoccimarro}}(2006{\natexlab{a}})}]{2006PhRvD..73f3519C}
\bibinfo{author}{\bibfnamefont{M.}~\bibnamefont{{Crocce}}} \bibnamefont{and}
  \bibinfo{author}{\bibfnamefont{R.}~\bibnamefont{{Scoccimarro}}},
  \bibinfo{journal}{\prd} \textbf{\bibinfo{volume}{73}},
  \bibinfo{pages}{063519} (\bibinfo{year}{2006}{\natexlab{a}}),
  \eprint{astro-ph/0509418}.

\bibitem[{\citenamefont{{Valageas}}(2007{\natexlab{a}})}]{2007A&A...465..725V}
\bibinfo{author}{\bibfnamefont{P.}~\bibnamefont{{Valageas}}},
  \bibinfo{journal}{\aap} \textbf{\bibinfo{volume}{465}}, \bibinfo{pages}{725}
  (\bibinfo{year}{2007}{\natexlab{a}}), \eprint{arXiv:astro-ph/0611849}.

\bibitem[{\citenamefont{{Matarrese} and
  {Pietroni}}(2007)}]{2007astro.ph..3563M}
\bibinfo{author}{\bibfnamefont{S.}~\bibnamefont{{Matarrese}}} \bibnamefont{and}
  \bibinfo{author}{\bibfnamefont{M.}~\bibnamefont{{Pietroni}}},
  \bibinfo{journal}{ArXiv Astrophysics e-prints}  (\bibinfo{year}{2007}),
  \eprint{astro-ph/0703563}.

\bibitem[{\citenamefont{{McDonald}}(2007)}]{2007PhRvD..75d3514M}
\bibinfo{author}{\bibfnamefont{P.}~\bibnamefont{{McDonald}}},
  \bibinfo{journal}{\prd} \textbf{\bibinfo{volume}{75}},
  \bibinfo{pages}{043514} (\bibinfo{year}{2007}),
  \eprint{arXiv:astro-ph/0606028}.

\bibitem[{\citenamefont{{Valageas}}(2007{\natexlab{b}})}]{2007arXiv0711.3407V}
\bibinfo{author}{\bibfnamefont{P.}~\bibnamefont{{Valageas}}},
  \bibinfo{journal}{ArXiv e-prints} \textbf{\bibinfo{volume}{711}}
  (\bibinfo{year}{2007}{\natexlab{b}}), \eprint{0711.3407}.

\bibitem[{\citenamefont{{Crocce} and
  {Scoccimarro}}(2006{\natexlab{b}})}]{2006PhRvD..73f3520C}
\bibinfo{author}{\bibfnamefont{M.}~\bibnamefont{{Crocce}}} \bibnamefont{and}
  \bibinfo{author}{\bibfnamefont{R.}~\bibnamefont{{Scoccimarro}}},
  \bibinfo{journal}{\prd} \textbf{\bibinfo{volume}{73}},
  \bibinfo{pages}{063520} (\bibinfo{year}{2006}{\natexlab{b}}),
  \eprint{astro-ph/0509419}.

\bibitem[{\citenamefont{{Crocce} and
  {Scoccimarro}}(2007)}]{2007arXiv0704.2783C}
\bibinfo{author}{\bibfnamefont{M.}~\bibnamefont{{Crocce}}} \bibnamefont{and}
  \bibinfo{author}{\bibfnamefont{R.}~\bibnamefont{{Scoccimarro}}},
  \bibinfo{journal}{ArXiv e-prints} \textbf{\bibinfo{volume}{704}}
  (\bibinfo{year}{2007}), \eprint{0704.2783}.

\bibitem[{\citenamefont{{Matsubara}}(2007)}]{2007arXiv0711.2521M}
\bibinfo{author}{\bibfnamefont{T.}~\bibnamefont{{Matsubara}}},
  \bibinfo{journal}{ArXiv e-prints} \textbf{\bibinfo{volume}{711}}
  (\bibinfo{year}{2007}), \eprint{0711.2521}.

\bibitem[{\citenamefont{{Valageas}}(2007{\natexlab{c}})}]{2007A&A..476..31V}
\bibinfo{author}{\bibfnamefont{P.}~\bibnamefont{{Valageas}}},
  \bibinfo{journal}{\aap} \textbf{\bibinfo{volume}{476}}, \bibinfo{pages}{31}
  (\bibinfo{year}{2007}{\natexlab{c}}), \eprint{arXiv/0706.2593}.

\bibitem[{\citenamefont{{Pichon} and {Bernardeau}}(1999)}]{1999A&A...343..663P}
\bibinfo{author}{\bibfnamefont{C.}~\bibnamefont{{Pichon}}} \bibnamefont{and}
  \bibinfo{author}{\bibfnamefont{F.}~\bibnamefont{{Bernardeau}}},
  \bibinfo{journal}{\aap} \textbf{\bibinfo{volume}{343}}, \bibinfo{pages}{663}
  (\bibinfo{year}{1999}), \eprint{arXiv:astro-ph/9902142}.

\bibitem[{\citenamefont{{Bernardeau}}(1994)}]{1994ApJ...427...51B}
\bibinfo{author}{\bibfnamefont{F.}~\bibnamefont{{Bernardeau}}},
  \bibinfo{journal}{\apj} \textbf{\bibinfo{volume}{427}}, \bibinfo{pages}{51}
  (\bibinfo{year}{1994}), \eprint{arXiv:astro-ph/9311066}.

\bibitem[{\citenamefont{{Bernardeau}}(2007)}]{BernardeauBook}
\bibinfo{author}{\bibfnamefont{F.}~\bibnamefont{{Bernardeau}}},
  \emph{\bibinfo{title}{Cosmologie, des fondements th{\'e}oriques aux
  observations}} (\bibinfo{publisher}{Editions du CNRS et EDP Sciences},
  \bibinfo{year}{2007}).

\bibitem[{\citenamefont{{Valageas}}(2004)}]{2004A&A...421...23V}
\bibinfo{author}{\bibfnamefont{P.}~\bibnamefont{{Valageas}}},
  \bibinfo{journal}{\aap} \textbf{\bibinfo{volume}{421}}, \bibinfo{pages}{23}
  (\bibinfo{year}{2004}), \eprint{arXiv:astro-ph/0307008}.

\end{thebibliography}

\end{document}